\documentclass[reprint,prb,superscriptaddress,amsmath,amssymb,aps,
]{revtex4-2}

\usepackage{bm}%
\usepackage{wasysym}
\usepackage{amssymb}
\usepackage{bbold}
\usepackage[colorlinks=true,linkcolor=blue]{hyperref}%
\expandafter\ifx\csname package@font\endcsname\relax\else
 \expandafter\expandafter
 \expandafter\usepackage
 \expandafter\expandafter
 \expandafter{\csname package@font\endcsname}%
\fi
\usepackage{graphicx}
\usepackage{dcolumn}

\begin{document}

\title{Spin Hall effect in electronic Lévy glasses: Enhanced spin current generation in the superdiffusive regime}

\author{Diego B. Fonseca}
\affiliation{Departamento de F\'{\i}sica, Centro de Ciências Exatas e da Natureza, Universidade Federal de Pernambuco, Recife - PE, 50670-901, Brazil}

\author{Luiz Felipe C. Pereira}
\email{luiz.cpereira@ufrpe.br}
\affiliation{Departamento de F\'{\i}sica, Centro de Ciências Exatas e da Natureza, Universidade Federal de Pernambuco, Recife - PE, 50670-901, Brazil}

\author{Anderson L. R. Barbosa}
\email{anderson.barbosa@ufrpe.br}
\affiliation{Departamento de F\'{\i}sica, Universidade Federal Rural de Pernambuco, Recife - PE, 52171-900, Brazil}

\begin{abstract}
In spintronics, both electronic charge and spin are used to process and store information. 
Generation, manipulation, and detection of spin currents are essential for the development of next-generation spintronic technologies.
Here, we investigate the spin Hall effect in electronic Lévy glasses composed of graphene ribbons with randomly distributed circular regions of high spin-orbit coupling. 
These systems exhibit two transport regimes that can be tuned by adjusting the Fermi energy. 
The superdiffusive regime is characterized by low Fermi energy, low resistivity, and low magnetoresistivity, resulting in a long spin diffusion length, in contrast to the diffusive regime.
Employing the Landauer-Büttiker approach in conjunction with numerically exact tight-binding simulations, we compute spin-resolved transmission coefficients to assess the spin Hall current and the spin Hall angle as functions of Fermi energy, spin-orbit coupling strength, and on-site electrostatic potential. 
Our findings reveal that, in the superdiffusive regime, a low charge current can be converted into a large spin Hall current, whereas in the diffusive regime, the same charge current generates a modest spin Hall current. 
Moreover, we observe that the spin Hall angle can reach 30\% in the superdiffusive regime, whereas in the diffusive regime it is only 5\%. 
These results demonstrate that electronic Lévy glasses provide a versatile platform for controlling spin transport and optimizing the spin Hall effect for spintronic applications.
\end{abstract}

\maketitle

\section{Introduction} 

Efficient generation, manipulation, and detection of spin currents are key topics of interest in spintronics, as they are essential for next-generation spintronic technologies [\onlinecite{RevModPhys.92.021003,RevModPhys.91.035004, RevModPhys.76.323, saitoh2017spin, doi:10.1021/acs.nanolett.3c00687,Perkins_2024,Choe2015,Victor2025,PhysRevApplied.10.031001,KIM2023155352,Zhou_2025}]. A central focus of this research is the spin Hall effect (SHE), a relativistic phenomenon that allows the conversion of a longitudinal charge current into a transverse spin Hall current through intrinsic or extrinsic spin-orbit coupling (SOC) [\onlinecite{jungwirth2012spin, PhysRevLett.83.1834, PhysRevLett.112.066601, PhysRevLett.85.393, dyakonov1971possibility, PhysRevB.94.134202,sahoo2025rashbainducedspinhallresponse}]. The development of spintronic devices that enhance the efficiency of this charge-to-spin conversion, measured by the spin Hall angle (SHA), without significantly reducing the spin diffusion length remains a primary objective in both fundamental condensed matter physics and materials engineering [\onlinecite{k5tg-xm6q,c5x3-yqf8}].

Graphene has emerged as a remarkable material for spin transport due to its long spin diffusion length [\onlinecite{PhysRevB.92.201410,doi:10.1021/acs.nanolett.6b00497,PhysRevLett.117.147201,doi:10.1021/acs.nanolett.8b04368}]. 
Also, due to its two-dimensional character, some of its electronic  properties can be tuned by an external gate voltage, underscoring its potential for both electronic and spintronic applications [\onlinecite{k5tg-xm6q,PhysRevLett.127.047202,doi:10.1021/acsaelm.5c00754,ramo,Mendes_2019,dueas2025optimalspinchargeinterconversiongraphene}]. 
However, its weak intrinsic SOC significantly limits its ability to enable charge-to-spin conversion, which is crucial for spintronic devices. 
To address this limitation, various strategies have been developed to induce an extrinsic SOC, such as chemical functionalization and adatom deposition [\onlinecite{PhysRevLett.110.246602, PhysRevLett.103.026804}]. 
Alternatively, proximity-induced SOC due to the presence of heavy metals [\onlinecite{doi:10.1021/acs.nanolett.3c00687,k5tg-xm6q}] and transition metal dichalcogenides [\onlinecite{PhysRevB.92.155403, PhysRevB.93.155104, PhysRevLett.119.146401,garcia2018spin,Yang_2024,https://doi.org/10.1002/adfm.202404872,PhysRevApplied.19.014053,PhysRevB.109.L241404,PhysRevResearch.3.033137,PhysRevB.103.L081111,doi:10.1021/acsami.4c08539}] has also been explored. 
The benefit of using proximity-induced methods lies in their ability to preserve graphene's electronic mobility and linear dispersion while enhancing spin-orbit effects [\onlinecite{doi:10.1021/acs.nanolett.8b04368,doi:10.1021/acs.nanolett.9b01611,PhysRevLett.119.196801,Ben_tez_2020,doi:10.1021/acsnano.0c01037,PhysRevLett.124.236803,Khokhriakov_2020}]. 

In this context, an electronic Lévy glass appears as a promising spintronic device, as it can induce a superdiffusive electronic transport regime in graphene rather than a standard diffusive one [\onlinecite{PhysRevB.107.155432}]. 
The superdiffusive regime is characterized by a low Fermi energy, low resistivity, and low magnetoresistivity, resulting in a long spin diffusion length. 
In contrast, the diffusive regime has a high Fermi energy, high resistivity, and a high magnetoresistivity, resulting in a short spin diffusion length [\onlinecite{PhysRevB.110.075421}]. 
Thus, the superdiffusive regime allows for more efficient charge-to-spin conversion than the diffusive one, as our results  demonstrate. 

The electronic Lévy glass consists of a graphene nanoribbon with randomly distributed circular regions, whose radii follow a power-law distribution. 
These circular regions are due to the presence of materials, such as heavy metals and transition metal dichalcogenides, which in contact with graphene form a bilayer structure, and can induce an on-site electrostatic potential [\onlinecite{Caridad2016,Christopher2016,https://doi.org/10.1002/pssb.201552119,PhysRevB.99.155432}] or enhance extrinsic SOC [\onlinecite{PhysRevLett.110.246602, PhysRevLett.103.026804,doi:10.1021/acs.nanolett.3c00687,k5tg-xm6q,PhysRevB.92.155403, PhysRevB.93.155104, PhysRevLett.119.146401,garcia2018spin,Yang_2024,https://doi.org/10.1002/adfm.202404872,PhysRevApplied.19.014053,PhysRevB.109.L241404,PhysRevResearch.3.033137,PhysRevB.103.L081111,doi:10.1021/acsami.4c08539}] in graphene. 
We have shown in Ref. [\onlinecite{PhysRevB.110.075421}] that an electronic Lévy glass connected to two leads exhibits finite spin polarization in the superdiffusive regime, while the polarization is nearly nonexistent in the diffusive regime. 
These findings motivated us to explore the SHE in the superdiffusive regime.

\begin{figure}
    \centering
    \includegraphics[width=\linewidth]{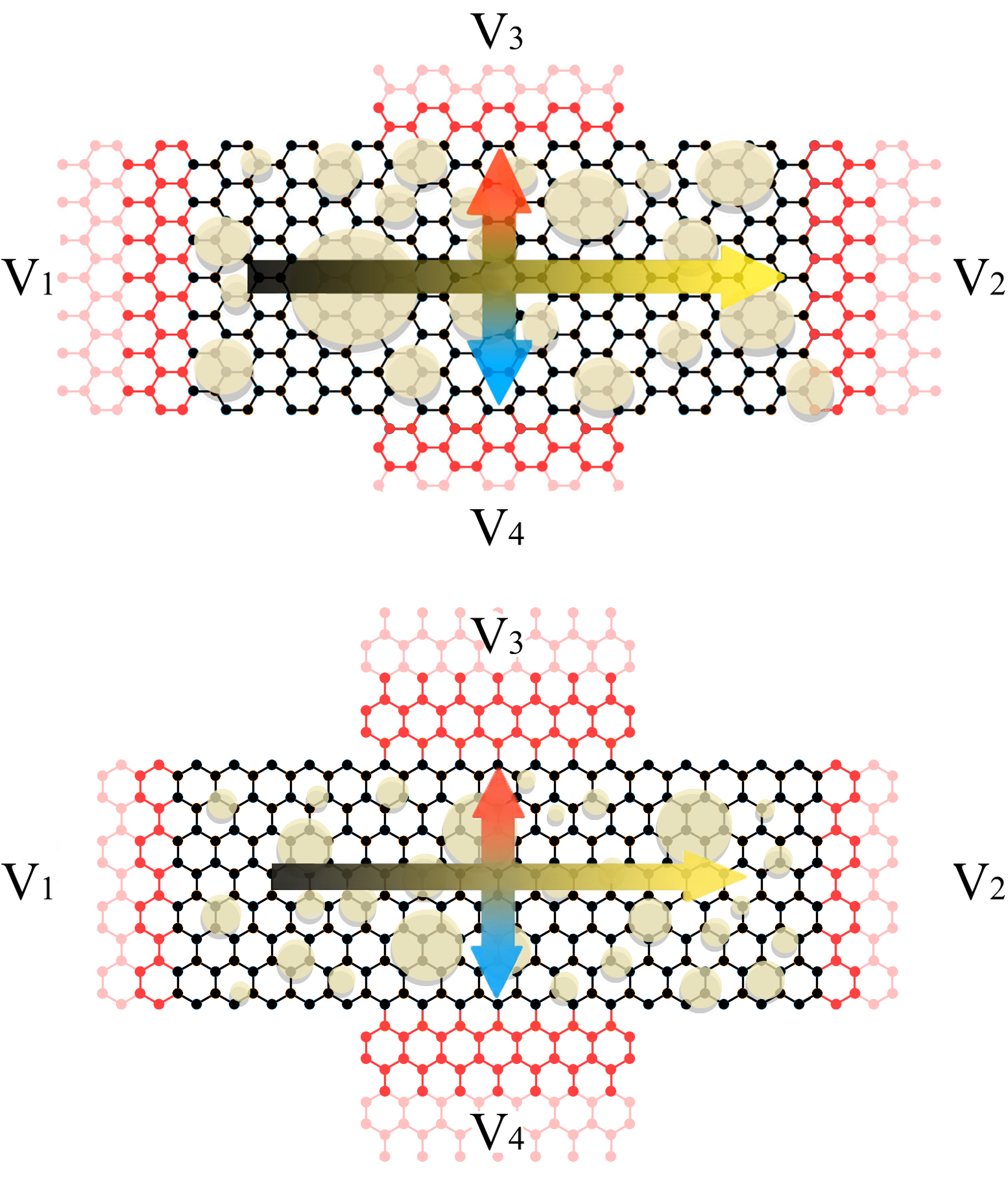}
    \caption{Illustration of (up) AGNR and (down) ZGNR (black sites) connected to four semi-infinite leads (red sites). Circular regions represent the graphene proximity-coupled to a high-SOC material. A charge current (longitudinal arrow) gives rise to spin Hall current (transversal arrows). $V_i$ is the potential in each lead. }
    \label{fig:4terminal_ELG}
\end{figure}

In this work, we investigate the SHE in an electronic Lévy glass embedded in a structure with enhanced extrinsic SOC and connected to four terminals, as shown in Fig.~\ref{fig:4terminal_ELG}. We employ the Landauer-Büttiker formulism to conduct numerically exact tight-binding calculations of the longitudinal charge current, the transverse spin Hall current, and SHA as functions of the Fermi energy and the SOC strength.
Our findings indicate that in the superdiffusive regime (at low Fermi energies), a small longitudinal charge current can be converted into a large transverse spin Hall current, whereas in the diffusive regime (at high Fermi energies), the same charge current generates only a modest spin Hall current.
Additionally, we found that the SHA can reach values as high as $30\%$ in the superdiffusive regime, whereas in the diffusive regime it only reaches $5\%$ for strong SOC. 
The latter is consistent with previously reported values in the literature [\onlinecite{k5tg-xm6q,Safeer_2020,Yang_2023,PhysRevB.109.L241404}].
Furthermore, we demonstrate that applying a local electrostatic potential to the SOC clusters can further enhance this efficiency. 
These findings position electronic Lévy glasses as a flexible platform for controlling charge-to-spin conversion through disorder engineering, offering valuable insights for the design of efficient and robust spintronic materials and devices.

\section{Methodology} \label{sec:method}

We investigate the SHE in both superdiffusive and diffusive transport regimes using an electronic Lévy glass device connected to four semi-infinite leads with the effect of proximity-induced SOC, as schematically illustrated in Fig. \ref{fig:4terminal_ELG}.
The device consists of an armchair (AGNR) or a zigzag (ZGNR) graphene nanoribbon with circular regions that emulate proximity coupling to a material with strong SOC. 
Those circular regions introduce a certain degree of disorder, since they behave as randomly distributed spin-orbit clusters due to the Bychkov-Rashba SOC.

The electronic Lévy glass is designed to display anomalous transport characteristics through a specific disorder geometry. 
In this setup, cluster centers are randomly distributed within the nanoribbon dimensions (length $L$ and width $W$) while adhering to strict non-overlap conditions. 
The radii of these circles, denoted as \(R\), follow a power-law distribution given by \(P(R) \propto R^{-(\beta + 1)}\), with \(\beta\) set to \(1.22\). 
The distribution has explicit lower and upper cutoffs: the minimum radius is determined by the lattice scale, set at \(R_{\mathrm{min}} = a\), where \(a= 2.49\) \AA~ is the graphene lattice constant, and the maximum radius is constrained by geometry, defined as \(R_{\mathrm{max}} = W/8\).  
The non-overlap condition, combined with the sampling procedure, results in a typical area coverage of the spin-orbit clusters of approximately \(42.24\% \pm 0.03\) for the device sizes considered in the study. 
This heavy-tailed radii distribution enables transition between superdiffusive and diffusive transport regimes, as detailed in Refs. [\onlinecite{PhysRevB.107.155432, PhysRevB.110.075421}].
The AGNR and ZGNR electronic Lévy glasses used in our numerical experiments have length $L = 1050a$ and width $W \approx 50a$ for AGNR, and $L = 1050a$ and $W \approx 50a$ for ZGNR. 

\subsection{Landauer-Büttiker formalism}

In order to characterize the SHE in our four-terminal electronic Lévy glass device, we employ the Landauer-Büttiker formalism [\onlinecite{PhysRevB.72.075361,PhysRevLett.98.196601,PhysRevB.102.041107,10.1063/5.0107212,PhysRevB.108.245105,PhysRevB.110.085412,barbosa2025extrinsicorbitalhalleffect}]. 
The spin-resolved current through the $i$th electrode is given by
\begin{equation}
    I^{S_\eta}_{i,\eta} = \frac{e^2}{h}\sum_{j} \tau_{ij,\eta}^{S_\eta} \left( V_i - V_j \right),\label{IOS}
\end{equation} 
where \(S_\eta\) are spin states [ \(\uparrow\) or \(\downarrow\) ], with $\eta = \{0,x,y,z\}$. 
The spin transmission coefficient $\tau_{ij}^{S_\eta}$ is calculated from the scattering matrix \(\mathcal{S}\) as
\begin{equation}
\tau_{ij}^{S_\eta} = \mathbf{Tr} \left[ (\mathcal{S}_{ij})^\dagger \mathcal{P}^{S_\eta}_{\eta} \mathcal{S}_{ij} \right], \quad \mathcal{S} = \begin{bmatrix} r_{11} & t_{12} & t_{13} & t_{14} \\ t_{21} & r_{22} & t_{23} & t_{24} \\ t_{31} & t_{32} & r_{33} & t_{34} \\ t_{41} & t_{42} & t_{43} & r_{44} \end{bmatrix}.
\end{equation}
The matrix $\mathcal{P}^{S_\eta}_{\eta} = \mathbb{1}_N \otimes \sigma^\eta$ is a projector, where $\mathbb{1}_N$ is an identity matrix with dimension $N\times N$. 
The dimensionless integer $N$ is the number of propagating modes in the leads, proportional to the lead width ($W$) and the Fermi vector ($k_F$) through the equation $N = k_F W/\pi$.  
The matrices \(\sigma^\eta\), with $\eta = \{0,x,y,z\}$, are spin matrices, and $\eta=0$ refers to the identity matrix in spin subspace. 
Thus, by setting either $\eta=0$ or $\eta=\{x,y,z\}$, charge and spin can be respectively addressed. The block matrices \(r_{ij}\) and \(t_{ij}\) represent the reflection and transmission components of the scattering matrix \(\mathcal{S}\), respectively.

The pure transversal spin Hall current $I^{S_z}_{i,z}=\frac{\hbar}{e} \left( I^{\uparrow}_{i}-I^{\downarrow}_{i}\right)$,  where $i=3,4$ can be obtained by assuming that the charge current vanishes in the transverse leads, $I^{c}_{i,0}=I^{\uparrow}_i+I^{\downarrow}_i=0$, while the charge current is conserved in the longitudinal leads, thus $I^c_{1,0}=-I^c_{2,0}=I^c$ [\onlinecite{PhysRevB.72.075361,PhysRevLett.98.196601}]. By applying these conditions to Eq.~(\ref{IOS}), we obtain
\begin{eqnarray}
I^{S_\eta}_{i,\eta} = \frac{e}{2\pi}\left[\left(\tau_{i2,\eta}^{S_\eta}-\tau_{i1,\eta}^{S_\eta}\right)\frac{V}{2}
- \left(\tau_{i3,\eta}^{S_\eta}V_3 + \tau_{i4,\eta}^{S_\eta}V_4\right)\right],
\label{Is}
\end{eqnarray}
for $i=3,4$, where $V$ is a constant potential difference between longitudinal leads, and $V_{3,4}$ is the transversal lead voltage.
The spin Hall angle is defined as the ratio between the transversal spin Hall current and the longitudinal charge current as follows

\begin{equation}
    \theta_\mathrm{SH} = \frac{e}{\hbar} \frac{I^{S_z}}{I^c}.
\end{equation}

\subsection{Tight-binding Model}

We perform numerical calculations of the transmission coefficients using KWANT [\onlinecite{kwant}], which implements a Green's function-based algorithm within the tight-binding framework. 
The Hamiltonian of graphene with  proximity-induced SOC is given by [\onlinecite{PhysRevB.98.045407,PhysRevB.103.L081111,PhysRevB.97.085413,PhysRevB.111.035411}]

\begin{equation}
    \hat{H} = -t\sum_{\left \langle i,j \right \rangle , \sigma}  c_{i\sigma}^\dagger c_{j\sigma}  
    - \sum_{\langle i,j \rangle, \sigma, \sigma^\prime } \imath \lambda_{ij} \, c_{i\sigma}^{\dagger} \left(\left[\textbf{s}\right]_{\sigma \sigma^\prime}\times \hat{\textbf{r}}_{ij}\right)_z c_{j\sigma^\prime}\,
    \label{H}
\end{equation}
where $i$ and $j$ enumerate all lattice sites, $\left \langle i,j \right \rangle$ denotes first nearest neighbors, $c_{i,\sigma}$ ($c_{i,\sigma}^\dagger$) are annihilation (creation) operators for electrons at site $i$ with spin $\sigma = \uparrow,\downarrow$, $\textbf{s}$ is the vector of Pauli matrices, and $\hat{\textbf{r}}_{ij}$ is the unit vector along the bond connecting sites $i$ and $j$.
The first term represents the nearest-neighbor hopping with $t = 2.7$ eV [\onlinecite{RevModPhys.81.109}]. The second term describes the Bychkov-Rashba SOC induced by proximity effects, which explicitly breaks inversion symmetry ($\vec{z} \rightarrow -\vec{z}$). The Rashba coupling $\lambda_{ij}$ takes the value $\lambda$ when both sites $i$ and $j$ reside within the same spin-orbit cluster, and vanishes otherwise.

It is important to consider that depositing graphene onto a substrate can introduce various effects, including electrostatic potential and strain, in addition to enhancing the SOC [\onlinecite{RevModPhys.91.035004,Perkins_2024,PhysRevB.110.075421}]. 
Therefore, to account for these potential defects and understand their impact on the SHE in the Lévy glass device, we include an on-site electrostatic potential in our model. 
Specifically, we add the following term to Eq.~\ref{H}
$$\hat{H}_{\text{on-site}} = \sum_{i,\sigma} U_i c_{i\sigma}^\dagger c_{i\sigma}$$
where \( U_i \) is set to zero for sites outside the cluster and equal to \( U \) for sites within the cluster. 

Overall, we consider extensive parameter sweeps across Fermi energy $E \in [0.0, 0.55t]$, Bychkov-Rashba SOC strength $\lambda \in [0.0, 0.79t]$, and on-site electrostatic potential \(U \in [0.0, 0.5t]\), with ensemble averaging over \(10^3\) different cluster configurations. 

\begin{figure*}
    \centering
\includegraphics[width=0.8\linewidth]{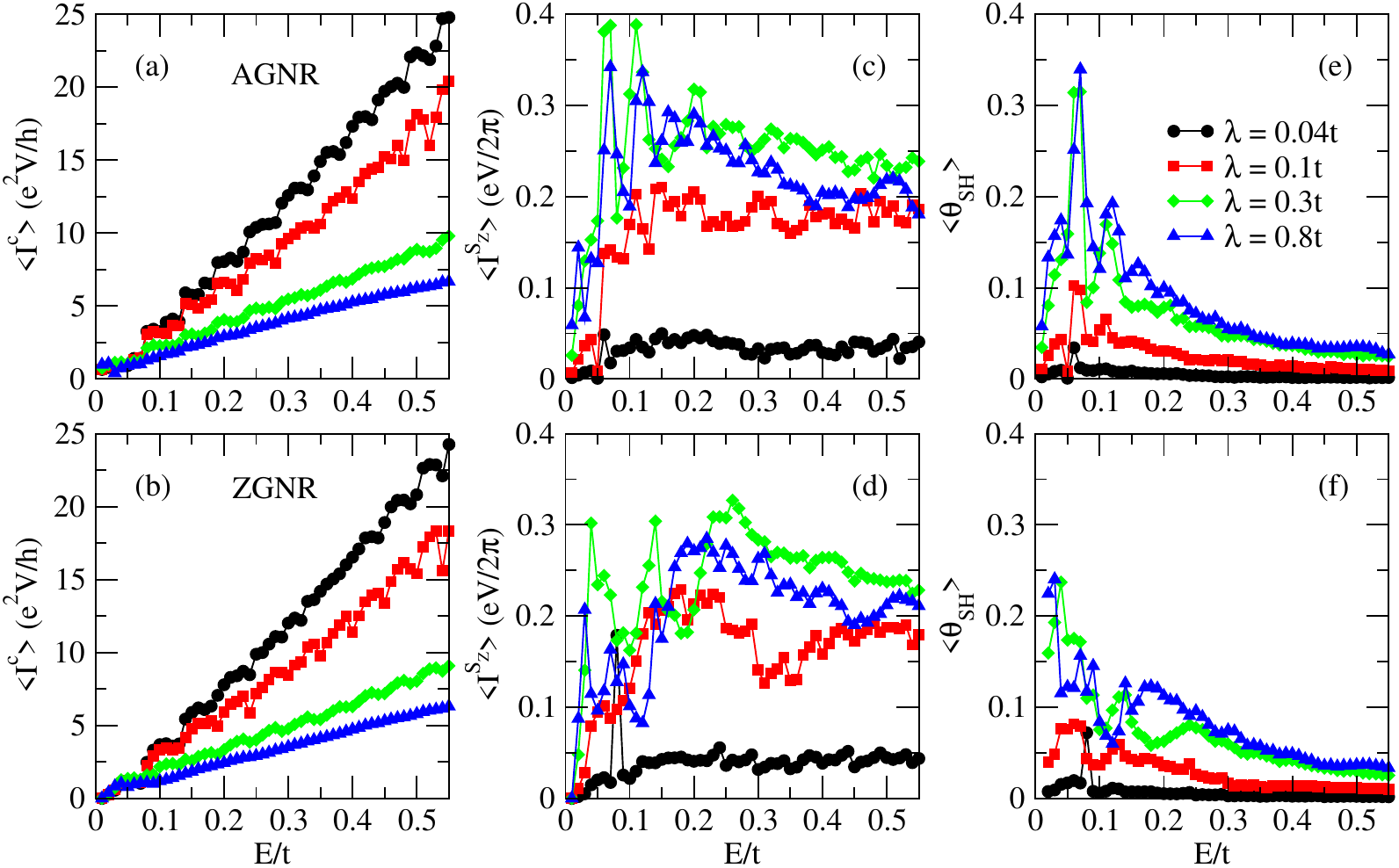}
    \caption{Averages of (a,b) longitudinal charge current, (c,d) transverse spin Hall current, and (e,f) spin Hall angle as functions of the Fermi energy. The top row (a, c, e) corresponds to AGNR results, while the bottom row (b, d, f) corresponds to ZGNR. Different symbols indicate different SOC strength: \(\lambda = 0.04t\) (black circles), \(\lambda = 0.1t\) (red squares), \(\lambda = 0.3t\) (green diamonds), and \(\lambda = 0.8t\) (blue triangles). Lines are guides to the eye. The averages are computed over \(10^3\) different samples.}
    \label{fig:Current_Angle_vs_Energy}
\end{figure*}

\section{Results and Discussion}\label{sec:discussion}

With the intent of investigating how the superdiffusive regime affects the SHE, in contrast to the diffusive one, we begin by analyzing the longitudinal charge current \( I^c \), the transverse spin Hall current \( I^{S_z} \), and the SHA \( \theta_\mathrm{SH} \) as functions of the Fermi energy \( E \).
We consider different values for the SOC strength while the on-site electrostatic potential is set to \(U=0\). 
Figures~\ref{fig:Current_Angle_vs_Energy}(a) and \ref{fig:Current_Angle_vs_Energy}(b) display the average longitudinal charge current for AGNR and ZGNR, respectively. 
As expected, the charge current for both types of ribbons increases with energy. 
However, the slope of the charge current decreases with increasing SOC strength, \(\lambda\), indicating that SOC clusters act as scattering centers, thereby generating disorder in the device, in agreement with [\onlinecite{PhysRevB.110.075421}].

Figures \ref{fig:Current_Angle_vs_Energy}(c) and \ref{fig:Current_Angle_vs_Energy}(d) show the average transverse spin Hall current as a function of energy for AGNR and ZGNR, respectively. 
In both cases, the curves exhibit a similar trend: the spin Hall current is nearly zero at \(E = 0\), increases rapidly at the superdiffusive regime (low energy, where $E<0.1t$), and then gradually decreases, reaching a saturation at the diffusive regime (high energy). 
For weak SOC $\lambda = 0.04t$, the spin Hall current saturates around $0.05\text{eV}/(2\pi)$.
In contrast, for stronger SOC values ($\lambda = 0.1t$, $0.3t$, and $0.8t$), the saturation value rises to approximately $0.2\text{eV}/(2\pi)$ for the diffusive regime, in agreement with [\onlinecite{PhysRevB.72.075361}]. 
On the other hand, in the superdiffusive regime, the spin Hall current can exceed $0.2\text{eV}/(2\pi)$ for strong SOC. 
Notably, in the superdiffusive regime, a large spin Hall current is generated by a much lower transverse charge current (in comparison with the the diffusive regime), yielding a higher charge-to-spin conversion efficiency.

The charge-to-spin conversion efficiency, represented by the SHA, is shown in Figs.~\ref{fig:Current_Angle_vs_Energy}(e) for AGNR and \ref{fig:Current_Angle_vs_Energy}(f) for ZGNR. 
As in the case of the spin Hall current, the results show a clear dependence on both $\lambda$ and the transport regime.
For weak SOC $\lambda = 0.04 t$, the efficiency is low, reaching a maximum value  $\left<\theta_\mathrm{SH}\right> \approx 5\%$ in the superdiffusive regime. 
At $\lambda = 0.1 t$, the maximum  value of $\left<\theta_\mathrm{SH}\right> \approx 10\%$ is also observed in the superdiffusive regime, followed by a decline as the system transitions into the diffusive regime. 
For $\lambda > 0.1 t$, the superdiffusive regime exhibits significantly higher efficiencies, attaining $\left<\theta_\mathrm{SH}\right> \approx 35\%$ for AGNR and $\left<\theta_\mathrm{SH}\right> \approx 25\%$ for ZGNR, although these values decrease with increasing energy. 
Remarkably, the response for $\lambda = 0.8 t$ is qualitatively similar to that for $\lambda = 0.3 t$, particularly at higher energies, suggesting a saturation effect with respect to $\lambda$.

\begin{figure*}
    \centering    \includegraphics[width=0.8\linewidth]{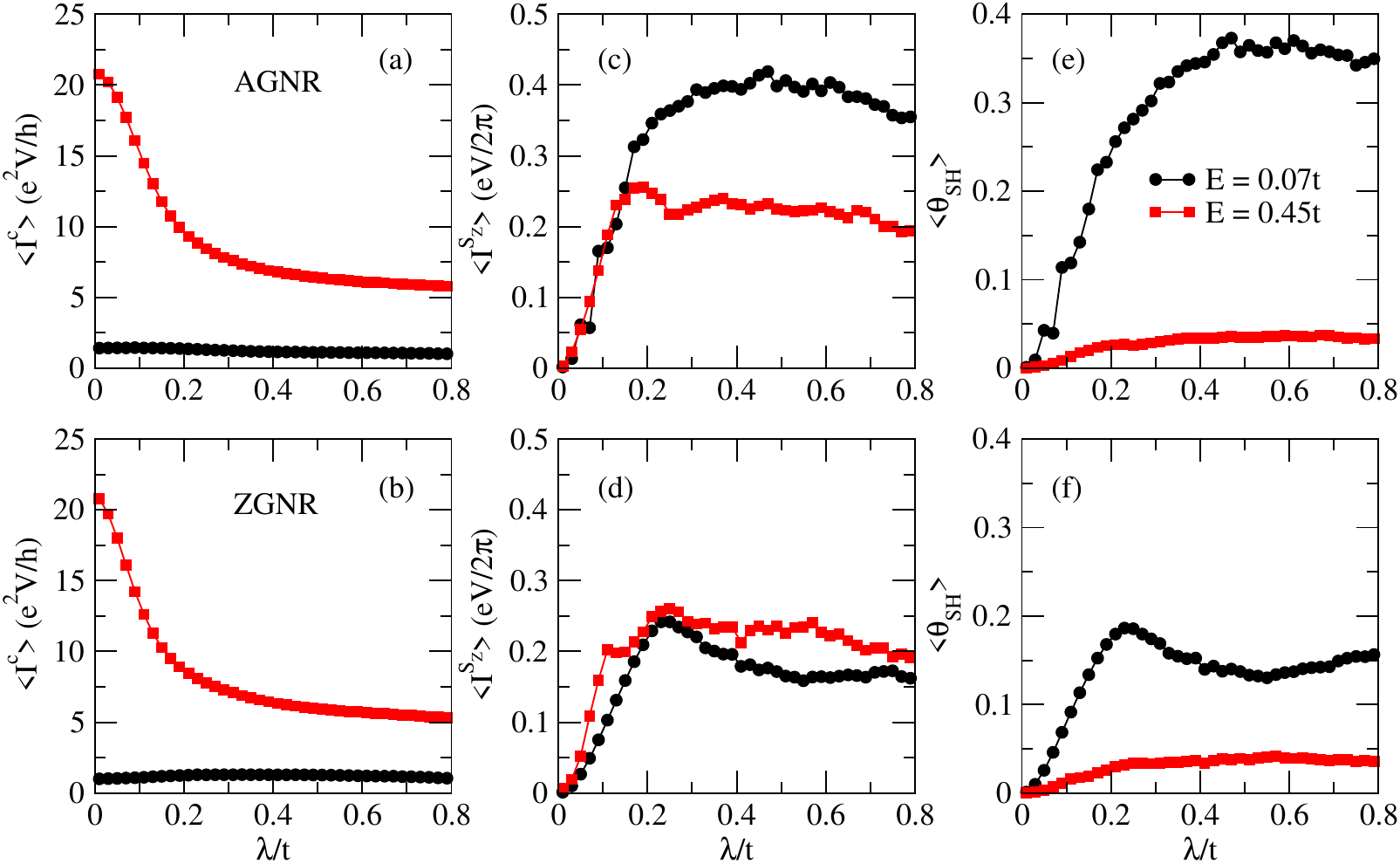}
    \caption{Averages of (a,b) longitudinal charge current, (c,d) transverse spin Hall current, and (e,f) spin Hall angle as functions of the Bychkov-Rashba spin-orbit coupling strength $\lambda$. The top row (a, c, e) corresponds to AGNR results, while the bottom row (b, d, f) corresponds to ZGNR. Different symbols indicate different regimes: superdiffusive \(E = 0.07t\) (black circles), diffusive \(E = 0.45t\) (red squares). Lines are guides to the eye. Each data point represents an average computed over \(10^3\) independent samples.}
    \label{fig:Current_Angle_vs_SOC}
\end{figure*}

Having established that charge-to-spin conversion efficiency is maximized in the superdiffusive regime, we now conduct a detailed comparison between the two transport regimes by fixing the energy at \(E = 0.07 t\) (superdiffusive) and \(E = 0.45 t\) (diffusive) while varying the strength of SOC. 
As shown in Fig.~\ref{fig:Current_Angle_vs_SOC}(a) and \ref{fig:Current_Angle_vs_SOC}(b), the average longitudinal charge current in the superdiffusive regime (black circles) remains constant at approximately \(2\text{e}^2\text{V}/h\) for all ranges of \(\lambda\), consistent with a low resistivity. 
This suggests that the charge current is invariant under SOC strength at both edge terminations in the super\text{diff}usive regime. 
In contrast, in the diffusive regime (red squares), the charge current experiences a significant suppression due to an increase in \(\lambda\), dropping from \(20\text{e}^2\text{V}/h\) to \(\approx 5\text{e}^2\text{V}/h\), indicating that the resistivity increases as \(\lambda\) increases. 

Regarding the average transversal spin Hall current, Fig.~\ref{fig:Current_Angle_vs_SOC}(c) shows significant differences between superdiffusive and diffusive regimes in AGNR. The spin Hall currents are initially equivalent and increase monotonically with the SOC strength \(\lambda\). For \(\lambda > 0.2t\), the spin Hall current reaches a saturation point of approximately \(0.2 \, \text{eV} / (2\pi)\) in the diffusive regime, which is consistent with the findings presented in Fig.~\ref{fig:Current_Angle_vs_Energy}(c). In contrast, in the superdiffusive regime, the spin Hall current saturates around \(0.35 \, \text{eV} / (2\pi)\), indicating an increase of about 70\% compared to the diffusive regime. Notably, the charge current in the superdiffusive regime is significantly lower than that in the diffusive regime across all SOC values. This result confirms the high efficiency of charge-to-spin conversion achieved through superdiffusion, when compared to the diffusive regime. 
In contrast, Fig.~\ref{fig:Current_Angle_vs_SOC}(d) shows that the behavior of the spin Hall current in both the superdiffusive and diffusive regimes for ZGNR is quite similar. In each regime, the spin Hall current increases steadily with the SOC strength, reaching a saturation point of approximately \(0.2 \, \text{eV} / (2\pi)\). 
Therefore, one can achieve the same spin Hall current in both regimes, but in the superdiffusive case the required charge current is significantly lower. 
This again highlights the advantages of working in the superdiffusive regime compared to the diffusive one.

In order to support our analysis of spin Hall currents, we focus on the SHA average. Figs.~\ref{fig:Current_Angle_vs_SOC}(e) and \ref{fig:Current_Angle_vs_SOC}(f) show a significant enhancement in the charge-to-spin conversion efficiency within the superdiffusive regime. For the AGNR depicted in Fig.~\ref{fig:Current_Angle_vs_SOC}(e), \(\left<\theta_\mathrm{SH}\right>\) increases with the SOC strength  before reaching saturation, achieving a maximum value of $35\%$ in the superdiffusive regime. In contrast, in the diffusive regime, the efficiency saturates at a value of less than \(5\%\), consistent with previously reported values in the literature [\onlinecite{k5tg-xm6q,Safeer_2020,Yang_2023,PhysRevB.109.L241404}]. A similar trend is observed for the ZGNR shown in Fig.~\ref{fig:Current_Angle_vs_SOC}(f): the SHA increases with $\lambda$ and then levels off, reaching a peak of about $20\%$ in the superdiffusive regime, while remaining capped at $5\%$ in the diffusive case. Interestingly, although the spin Hall currents for ZGNR are comparable across the two transport regimes (as shown in Fig.~\ref{fig:Current_Angle_vs_SOC}(d)), the conversion efficiency is significantly higher in the superdiffusive regime. This highlights the important role of Lévy-glass-induced superdiffusive transport in enhancing the spin Hall response.

\begin{figure*}
    \centering
    \includegraphics[width=\linewidth]{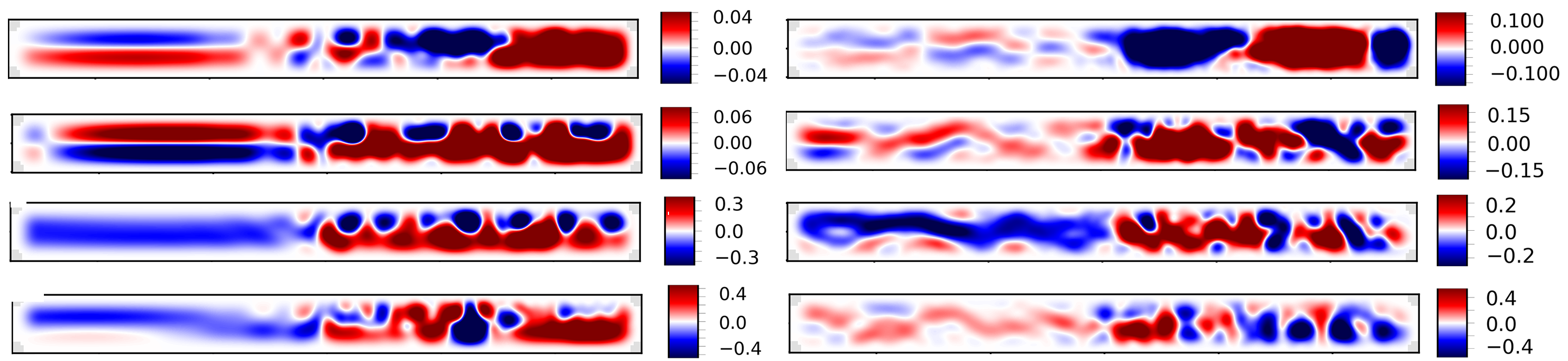}
    \caption{SDOS $\left \langle S_z \right \rangle$ for AGNR with $\lambda = 0.04t, 0.1t, 0.3t \text{ and } 0.8t$, top to bottom respectively. Left hand side column refers to a system in the superdiffusive regime $E=0.07t$, while the right hand side corresponds to the diffusive one $E = 0.45t$.}
    \label{fig:Sz_density}
\end{figure*}

To better understand the difference between superdiffusive and diffusive SHE in the electronic Lévy glass, we analyze the spin-polarized density of states (SDOS) \(\langle S_z \rangle\). Figure \ref{fig:Sz_density} displays the SDOS for AGNR at various SOC strengths: \(\lambda = 0.04t\), \(0.1t\), \(0.3t\), and \(0.8t\), arranged from top to bottom. The left hand side column illustrates the superdiffusive regime at \(E = 0.07t\), while the right hand side one shows the diffusive regime at \(E = 0.45t\).
The superdiffusive regime leads to more efficient spin-up and spin-down separation, particularly before the transverse terminals where $V_3$ and $V_4$ can be measured, as shown in Fig. \ref{fig:4terminal_ELG}.
Nonetheless, the spin separation is distributed throughout the device, as apparent on the left hand side of Fig. \ref{fig:Sz_density}. 
This feature of the superdiffusive regime leads to a decreased  magnetoresistivity across the device, thereby enhancing the transverse spin Hall current and, consequently, the efficiency of charge-to-spin conversion.  
In contrast, in the diffusive regime, spin-up and spin-down electrons are less segregated leading to an increased magnetoresistivity across the device and thereby reducing the efficiency of charge-to-spin conversion. 

\begin{figure}
    \centering
    \includegraphics[width=\linewidth]{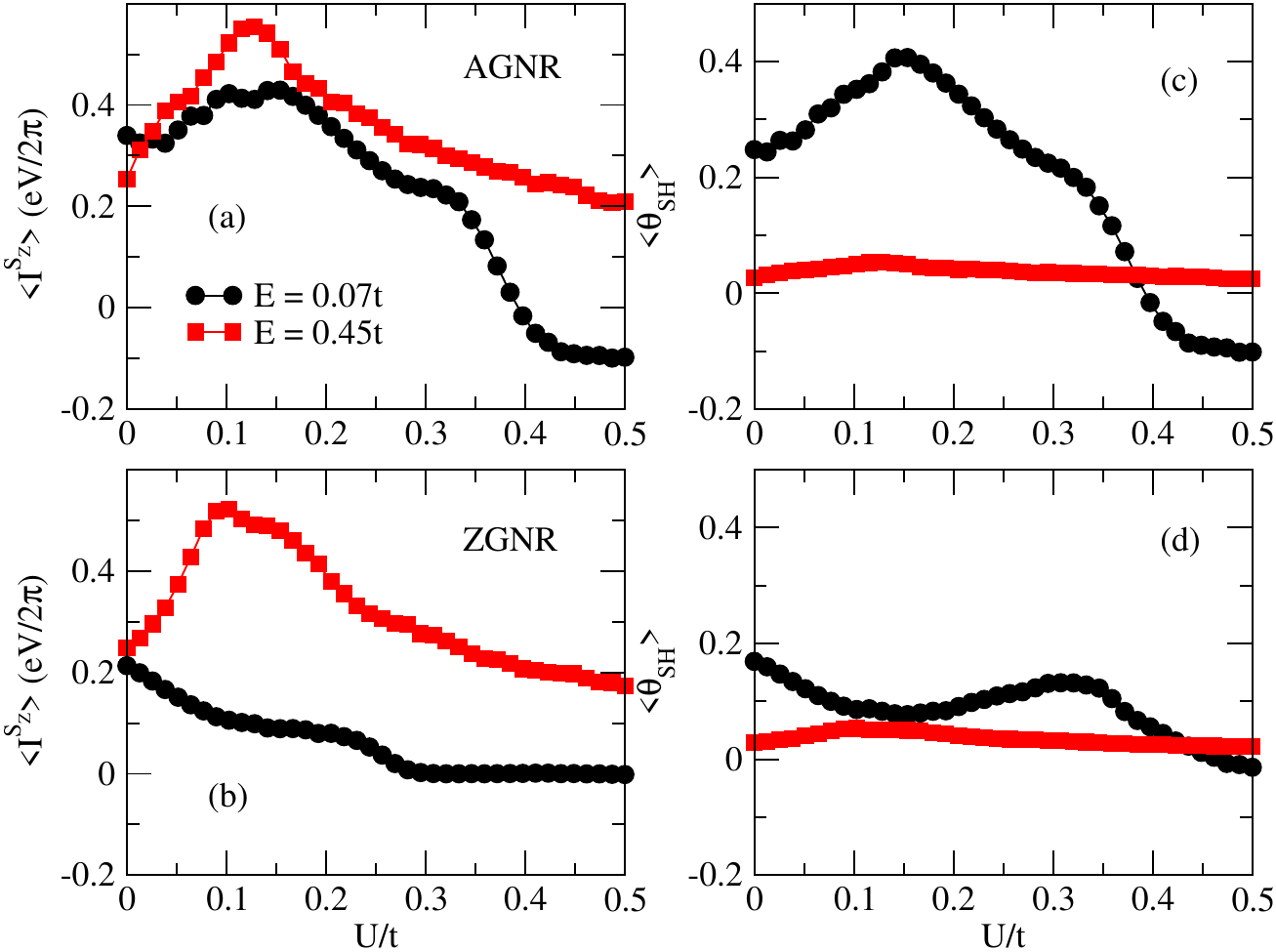}
    \caption{Average (a,b) transversal spin current and (c,d) SHA as functions of the on-site potential \(U\) at fixed SOC strength $\lambda = 0.2t$. The top row shows AGNR results, while the bottom row shows ZGNR. Different symbols indicate different regimes: superdiffusive \(E = 0.07t\) (black circles), diffusive \(E = 0.45t\) (red squares). Lines are guides to the eye. The averages are computed over \(10^3\) independent samples.}
    \label{fig:Angle_vs_Potential}
\end{figure}

In a final analysis, we calculate the transverse spin Hall current and the SHA as functions of the on-site electrostatic potential \(U\), as shown in Fig.~\ref{fig:Angle_vs_Potential}. 
While substrate interactions give rise to proximity-induced SOC, they also often introduce local electrostatic potentials.
Furthermore, previous studies have shown that weak electrostatic potentials can enhance SOC-induced spin polarization [\onlinecite{PhysRevB.110.075421}]. 
Figures~\ref{fig:Angle_vs_Potential}(a) and \ref{fig:Angle_vs_Potential}(b) show the average spin Hall current as a function of \(U\)  at \(\lambda=0.2t\) for AGNR and ZGNR, respectively. 
In both cases, we see that the spin Hall current in the diffusive regime (red squares) initially increases with the electrostatic potential, then decreases again to \(0.2 \, \text{eV} / (2\pi)\) for high values of \(U\). 
In contrast, the spin Hall current tends to be suppressed by the electrostatic potential in the superdiffusive regime (black circles). 
Thus, the spin Hall current in the diffusive regime is more sensitive to the on-site electrostatic potential, achieving a higher peak value than in the superdiffusive regime. 

Nonetheless, analyzing the behavior of the SHA, we find that the charge-to-spin conversion efficiency in the diffusive regime is not enhanced by the electrostatic potential, continuing to be around \(5\% \) independent of \(U\), as shown  in Figs.~\ref{fig:Angle_vs_Potential}(c) and \ref{fig:Angle_vs_Potential}(d).
This results implies that, even though there is an increase in spin Hall current induced by a weak electrostatic potential, it is not sufficient to increase the conversion efficiency significantly in the diffusive regime. 
In contrast, the SHA in the superdiffusive regime is strongly affected by the electrostatic potential, {and could be tuned with an increase in \(U\)}, as shown in Figs.~\ref{fig:Angle_vs_Potential}(c) and \ref{fig:Angle_vs_Potential}(d). In fact, the SHA can reach  \(40\%\) for AGNR and  \(20\%\) for ZGNR.

\section{Conclusions} \label{sec:conslusions}

In conclusion, we investigated the spin Hall effect in an electronic Lévy glass, composed of a graphene nanoribbon in the presence of randomly distributed spin-orbit clusters, whose radii follow a power-law distribution [\onlinecite{PhysRevB.107.155432, PhysRevB.110.075421}].
The presence of such clusters induce a superdiffusive dynamics on the electrons, in addition to the standard diffusive one, and each regime presents particular features in terms of spin transport.

We have shown that in the superdiffusive regime a
small longitudinal charge current can be converted into a large transverse spin Hall current, whereas in the diffusive regime, the same charge current generates only a modest spin Hall current.
Additionally, we have found that the SHA can reach 30\% in the superdiffusive regime, whereas in the diffusive regime it is only 5\% under strong spin-orbit coupling. 
More specifically, in the superdiffusive regime, the spin Hall current can increase by approximately 70\% relative to the diffusive regime.

Furthermore, we demonstrated that applying a local electrostatic potential to the spin-orbit clusters can further enhance this efficiency. 
These compelling results position electronic Lévy glasses as a powerful and adaptable platform for the controlled conversion of charge-to-spin current via disorder engineering. 
This insight is not only groundbreaking but is also essential for advancing the design of efficient and robust spintronic materials and devices, thereby shaping the future of the field.

Finally, it is important to discuss the possible experimental realization of electronic Lévy glasses. 
Ref. [\onlinecite{Caridad2016}] presented an experimental demonstration of an electronic analog of Mie scattering by utilizing a graphene superlattice as a conductor, which is integrated into a regular array of on-site electrostatic quantum dots. 
The experiment can be adapted to work as an electronic Lévy glass, in which the cluster radii follow a power-law distribution, as reported in [\onlinecite{PhysRevB.107.155432}]. 
Similarly, the experiments on proximity-induced SOC with heavy metals [\onlinecite{doi:10.1021/acs.nanolett.3c00687,k5tg-xm6q}] and transition metal dichalcogenides have also been explored [\onlinecite{PhysRevB.92.155403, PhysRevB.93.155104, PhysRevLett.119.146401, garcia2018spin, Yang_2024, https://doi.org/10.1002/adfm.202404872, PhysRevApplied.19.014053, PhysRevB.109.L241404, PhysRevResearch.3.033137, PhysRevB.103.L081111}]. 
These studies can be extended to electronic Lévy glass to enable the deposition of materials in which the cluster radii follow a power-law distribution. 
We believe that such extension could be achieved nowadays due to the current experimental precision on bilayer growth in spintronics devices [\onlinecite{k5tg-xm6q, Victor2025,doi:10.1021/acsami.4c08539}].

\section*{Acknowledgements}

DBF acknowledges a scholarship from Coordenação de Aperfeiçoamento de Pessoal de Nível Superior (CAPES, Grant 0041/2022).
LFCP acknowledges financial support from CAPES (Grant 0041/2022), Conselho Nacional de Desenvolvimento Cient\'{\i}fico e Tecnol\'ogico (CNPq, Grants  200296/2023-0, 371610/2023-0 INCT Materials Informatics, and 310262/2025-9), FACEPE (Grant APQ-1117-1.05/22).
ALRB acknowledges financial support from CNPq (Grants 302502/2025-4 and 406836/2022-1 INCT of Spintronics and Advanced Magnetic Nanostructures - SpinNanoMag).

\bibliography{ref.bib}

\begin{thebibliography}{68}%
\makeatletter
\providecommand \@ifxundefined [1]{%
 \@ifx{#1\undefined}
}%
\providecommand \@ifnum [1]{%
 \ifnum #1\expandafter \@firstoftwo
 \else \expandafter \@secondoftwo
 \fi
}%
\providecommand \@ifx [1]{%
 \ifx #1\expandafter \@firstoftwo
 \else \expandafter \@secondoftwo
 \fi
}%
\providecommand \natexlab [1]{#1}%
\providecommand \enquote  [1]{``#1''}%
\providecommand \bibnamefont  [1]{#1}%
\providecommand \bibfnamefont [1]{#1}%
\providecommand \citenamefont [1]{#1}%
\providecommand \href@noop [0]{\@secondoftwo}%
\providecommand \href [0]{\begingroup \@sanitize@url \@href}%
\providecommand \@href[1]{\@@startlink{#1}\@@href}%
\providecommand \@@href[1]{\endgroup#1\@@endlink}%
\providecommand \@sanitize@url [0]{\catcode `\\12\catcode `\$12\catcode
  `\&12\catcode `\#12\catcode `\^12\catcode `\_12\catcode `\%12\relax}%
\providecommand \@@startlink[1]{}%
\providecommand \@@endlink[0]{}%
\providecommand \url  [0]{\begingroup\@sanitize@url \@url }%
\providecommand \@url [1]{\endgroup\@href {#1}{\urlprefix }}%
\providecommand \urlprefix  [0]{URL }%
\providecommand \Eprint [0]{\href }%
\providecommand \doibase [0]{https://doi.org/}%
\providecommand \selectlanguage [0]{\@gobble}%
\providecommand \bibinfo  [0]{\@secondoftwo}%
\providecommand \bibfield  [0]{\@secondoftwo}%
\providecommand \translation [1]{[#1]}%
\providecommand \BibitemOpen [0]{}%
\providecommand \bibitemStop [0]{}%
\providecommand \bibitemNoStop [0]{.\EOS\space}%
\providecommand \EOS [0]{\spacefactor3000\relax}%
\providecommand \BibitemShut  [1]{\csname bibitem#1\endcsname}%
\let\auto@bib@innerbib\@empty
\bibitem [{\citenamefont {Avsar}\ \emph {et~al.}(2020)\citenamefont {Avsar},
  \citenamefont {Ochoa}, \citenamefont {Guinea}, \citenamefont {\"Ozyilmaz},
  \citenamefont {van Wees},\ and\ \citenamefont
  {Vera-Marun}}]{RevModPhys.92.021003}%
  \BibitemOpen
  \bibfield  {author} {\bibinfo {author} {\bibfnamefont {A.}~\bibnamefont
  {Avsar}}, \bibinfo {author} {\bibfnamefont {H.}~\bibnamefont {Ochoa}},
  \bibinfo {author} {\bibfnamefont {F.}~\bibnamefont {Guinea}}, \bibinfo
  {author} {\bibfnamefont {B.}~\bibnamefont {\"Ozyilmaz}}, \bibinfo {author}
  {\bibfnamefont {B.~J.}\ \bibnamefont {van Wees}},\ and\ \bibinfo {author}
  {\bibfnamefont {I.~J.}\ \bibnamefont {Vera-Marun}},\ }\bibfield  {title}
  {\bibinfo {title} {Colloquium: Spintronics in graphene and other
  two-dimensional materials},\ }\href
  {https://doi.org/10.1103/RevModPhys.92.021003} {\bibfield  {journal}
  {\bibinfo  {journal} {Rev. Mod. Phys.}\ }\textbf {\bibinfo {volume} {92}},\
  \bibinfo {pages} {021003} (\bibinfo {year} {2020})}\BibitemShut {NoStop}%
\bibitem [{\citenamefont {Manchon}\ \emph {et~al.}(2019)\citenamefont
  {Manchon}, \citenamefont {\ifmmode~\check{Z}\else \v{Z}\fi{}elezn\'y},
  \citenamefont {Miron}, \citenamefont {Jungwirth}, \citenamefont {Sinova},
  \citenamefont {Thiaville}, \citenamefont {Garello},\ and\ \citenamefont
  {Gambardella}}]{RevModPhys.91.035004}%
  \BibitemOpen
  \bibfield  {author} {\bibinfo {author} {\bibfnamefont {A.}~\bibnamefont
  {Manchon}}, \bibinfo {author} {\bibfnamefont {J.}~\bibnamefont
  {\ifmmode~\check{Z}\else \v{Z}\fi{}elezn\'y}}, \bibinfo {author}
  {\bibfnamefont {I.~M.}\ \bibnamefont {Miron}}, \bibinfo {author}
  {\bibfnamefont {T.}~\bibnamefont {Jungwirth}}, \bibinfo {author}
  {\bibfnamefont {J.}~\bibnamefont {Sinova}}, \bibinfo {author} {\bibfnamefont
  {A.}~\bibnamefont {Thiaville}}, \bibinfo {author} {\bibfnamefont
  {K.}~\bibnamefont {Garello}},\ and\ \bibinfo {author} {\bibfnamefont
  {P.}~\bibnamefont {Gambardella}},\ }\bibfield  {title} {\bibinfo {title}
  {Current-induced spin-orbit torques in ferromagnetic and antiferromagnetic
  systems},\ }\href {https://doi.org/10.1103/RevModPhys.91.035004} {\bibfield
  {journal} {\bibinfo  {journal} {Rev. Mod. Phys.}\ }\textbf {\bibinfo {volume}
  {91}},\ \bibinfo {pages} {035004} (\bibinfo {year} {2019})}\BibitemShut
  {NoStop}%
\bibitem [{\citenamefont {\ifmmode \check{Z}\else
  \v{Z}\fi{}uti\ifmmode~\acute{c}\else \'{c}\fi{}}\ \emph
  {et~al.}(2004)\citenamefont {\ifmmode \check{Z}\else
  \v{Z}\fi{}uti\ifmmode~\acute{c}\else \'{c}\fi{}}, \citenamefont {Fabian},\
  and\ \citenamefont {Das~Sarma}}]{RevModPhys.76.323}%
  \BibitemOpen
  \bibfield  {author} {\bibinfo {author} {\bibfnamefont {I.}~\bibnamefont
  {\ifmmode \check{Z}\else \v{Z}\fi{}uti\ifmmode~\acute{c}\else \'{c}\fi{}}},
  \bibinfo {author} {\bibfnamefont {J.}~\bibnamefont {Fabian}},\ and\ \bibinfo
  {author} {\bibfnamefont {S.}~\bibnamefont {Das~Sarma}},\ }\bibfield  {title}
  {\bibinfo {title} {Spintronics: Fundamentals and applications},\ }\href
  {https://doi.org/10.1103/RevModPhys.76.323} {\bibfield  {journal} {\bibinfo
  {journal} {Rev. Mod. Phys.}\ }\textbf {\bibinfo {volume} {76}},\ \bibinfo
  {pages} {323} (\bibinfo {year} {2004})}\BibitemShut {NoStop}%
\bibitem [{\citenamefont {Saitoh}\ \emph {et~al.}(2017)\citenamefont {Saitoh},
  \citenamefont {Maekawa}, \citenamefont {Valenzuela}, \citenamefont {Kimura},\
  and\ \citenamefont {Kimura}}]{saitoh2017spin}%
  \BibitemOpen
  \bibfield  {author} {\bibinfo {author} {\bibfnamefont {E.}~\bibnamefont
  {Saitoh}}, \bibinfo {author} {\bibfnamefont {S.}~\bibnamefont {Maekawa}},
  \bibinfo {author} {\bibfnamefont {S.}~\bibnamefont {Valenzuela}}, \bibinfo
  {author} {\bibfnamefont {K.}~\bibnamefont {Kimura}},\ and\ \bibinfo {author}
  {\bibfnamefont {T.}~\bibnamefont {Kimura}},\ }\href
  {https://books.google.com.br/books?id=qMm3swEACAAJ} {\emph {\bibinfo {title}
  {Spin Current}}},\ Series on semiconductor science and technology\ (\bibinfo
  {publisher} {Oxford University Press},\ \bibinfo {year} {2017})\BibitemShut
  {NoStop}%
\bibitem [{\citenamefont {Yang}\ \emph
  {et~al.}(2023{\natexlab{a}})\citenamefont {Yang}, \citenamefont {Ormaza},
  \citenamefont {Chi}, \citenamefont {Dolan}, \citenamefont {Ingla-Ayn{\'e}s},
  \citenamefont {Safeer}, \citenamefont {Herling}, \citenamefont {Ontoso},
  \citenamefont {Gobbi}, \citenamefont {Martín-García}, \citenamefont
  {Schiller}, \citenamefont {Hueso},\ and\ \citenamefont
  {Casanova}}]{doi:10.1021/acs.nanolett.3c00687}%
  \BibitemOpen
  \bibfield  {author} {\bibinfo {author} {\bibfnamefont {H.}~\bibnamefont
  {Yang}}, \bibinfo {author} {\bibfnamefont {M.}~\bibnamefont {Ormaza}},
  \bibinfo {author} {\bibfnamefont {Z.}~\bibnamefont {Chi}}, \bibinfo {author}
  {\bibfnamefont {E.}~\bibnamefont {Dolan}}, \bibinfo {author} {\bibfnamefont
  {J.}~\bibnamefont {Ingla-Ayn{\'e}s}}, \bibinfo {author} {\bibfnamefont
  {C.}~\bibnamefont {Safeer}}, \bibinfo {author} {\bibfnamefont
  {F.}~\bibnamefont {Herling}}, \bibinfo {author} {\bibfnamefont
  {N.}~\bibnamefont {Ontoso}}, \bibinfo {author} {\bibfnamefont
  {M.}~\bibnamefont {Gobbi}}, \bibinfo {author} {\bibfnamefont
  {B.}~\bibnamefont {Martín-García}}, \bibinfo {author} {\bibfnamefont
  {F.}~\bibnamefont {Schiller}}, \bibinfo {author} {\bibfnamefont {L.~E.}\
  \bibnamefont {Hueso}},\ and\ \bibinfo {author} {\bibfnamefont
  {F.}~\bibnamefont {Casanova}},\ }\bibfield  {title} {\bibinfo {title}
  {Gate-tunable spin hall effect in an all-light-element heterostructure:
  Graphene with copper oxide},\ }\href
  {https://doi.org/10.1021/acs.nanolett.3c00687} {\bibfield  {journal}
  {\bibinfo  {journal} {Nano Letters}\ }\textbf {\bibinfo {volume} {23}},\
  \bibinfo {pages} {4406} (\bibinfo {year} {2023}{\natexlab{a}})},\ \bibinfo
  {note} {pMID: 37140909},\ \Eprint
  {https://arxiv.org/abs/https://doi.org/10.1021/acs.nanolett.3c00687}
  {https://doi.org/10.1021/acs.nanolett.3c00687} \BibitemShut {NoStop}%
\bibitem [{\citenamefont {Perkins}\ and\ \citenamefont
  {Ferreira}(2024)}]{Perkins_2024}%
  \BibitemOpen
  \bibfield  {author} {\bibinfo {author} {\bibfnamefont {D.~T.}\ \bibnamefont
  {Perkins}}\ and\ \bibinfo {author} {\bibfnamefont {A.}~\bibnamefont
  {Ferreira}},\ }\bibinfo {title} {Spintronics in 2d graphene-based van der
  waals heterostructures},\ in\ \href
  {https://doi.org/10.1016/b978-0-323-90800-9.00203-1} {\emph {\bibinfo
  {booktitle} {Encyclopedia of Condensed Matter Physics}}}\ (\bibinfo
  {publisher} {Elsevier},\ \bibinfo {year} {2024})\ p.\ \bibinfo {pages}
  {205–222}\BibitemShut {NoStop}%
\bibitem [{\citenamefont {Choe}\ and\ \citenamefont {Chang}(2015)}]{Choe2015}%
  \BibitemOpen
  \bibfield  {author} {\bibinfo {author} {\bibfnamefont {D.-H.}\ \bibnamefont
  {Choe}}\ and\ \bibinfo {author} {\bibfnamefont {K.}~\bibnamefont {Chang}},\
  }\bibfield  {title} {\bibinfo {title} {Universal conductance fluctuation in
  two-dimensional topological insulators},\ }\href
  {https://doi.org/10.1038/srep10997} {\bibfield  {journal} {\bibinfo
  {journal} {Scientific reports}\ }\textbf {\bibinfo {volume} {5}},\ \bibinfo
  {pages} {10997} (\bibinfo {year} {2015})}\BibitemShut {NoStop}%
\bibitem [{\citenamefont {Victor}\ \emph {et~al.}(2025)\citenamefont {Victor},
  \citenamefont {Safeer}, \citenamefont {Marroquin}, \citenamefont {Costa},
  \citenamefont {Felix}, \citenamefont {Carozo}, \citenamefont {Sampaio},\ and\
  \citenamefont {García}}]{Victor2025}%
  \BibitemOpen
  \bibfield  {author} {\bibinfo {author} {\bibfnamefont {R.}~\bibnamefont
  {Victor}}, \bibinfo {author} {\bibfnamefont {S.}~\bibnamefont {Safeer}},
  \bibinfo {author} {\bibfnamefont {J.}~\bibnamefont {Marroquin}}, \bibinfo
  {author} {\bibfnamefont {M.}~\bibnamefont {Costa}}, \bibinfo {author}
  {\bibfnamefont {J.}~\bibnamefont {Felix}}, \bibinfo {author} {\bibfnamefont
  {V.}~\bibnamefont {Carozo}}, \bibinfo {author} {\bibfnamefont
  {L.}~\bibnamefont {Sampaio}},\ and\ \bibinfo {author} {\bibfnamefont
  {F.}~\bibnamefont {García}},\ }\bibfield  {title} {\bibinfo {title}
  {Disentangling edge and bulk spin-to-charge interconversion in mos2 monolayer
  flakes},\ }\href {https://doi.org/10.1038/s41467-025-58119-4} {\bibfield
  {journal} {\bibinfo  {journal} {Nature Communications}\ }\textbf {\bibinfo
  {volume} {16}} (\bibinfo {year} {2025})}\BibitemShut {NoStop}%
\bibitem [{\citenamefont {Zhu}\ \emph {et~al.}(2018)\citenamefont {Zhu},
  \citenamefont {Ralph},\ and\ \citenamefont
  {Buhrman}}]{PhysRevApplied.10.031001}%
  \BibitemOpen
  \bibfield  {author} {\bibinfo {author} {\bibfnamefont {L.}~\bibnamefont
  {Zhu}}, \bibinfo {author} {\bibfnamefont {D.~C.}\ \bibnamefont {Ralph}},\
  and\ \bibinfo {author} {\bibfnamefont {R.~A.}\ \bibnamefont {Buhrman}},\
  }\bibfield  {title} {\bibinfo {title} {Highly efficient spin-current
  generation by the spin hall effect in
  ${\mathrm{au}}_{1\ensuremath{-}x}{\mathrm{pt}}_{x}$},\ }\href
  {https://doi.org/10.1103/PhysRevApplied.10.031001} {\bibfield  {journal}
  {\bibinfo  {journal} {Phys. Rev. Appl.}\ }\textbf {\bibinfo {volume} {10}},\
  \bibinfo {pages} {031001} (\bibinfo {year} {2018})}\BibitemShut {NoStop}%
\bibitem [{\citenamefont {Kim}\ \emph {et~al.}(2023)\citenamefont {Kim},
  \citenamefont {Nguyen}, \citenamefont {{Won Kim}}, \citenamefont {{Hyeok
  Lee}}, \citenamefont {{In Yoon}}, \citenamefont {Rhim},\ and\ \citenamefont
  {{Keun Kim}}}]{KIM2023155352}%
  \BibitemOpen
  \bibfield  {author} {\bibinfo {author} {\bibfnamefont {T.}~\bibnamefont
  {Kim}}, \bibinfo {author} {\bibfnamefont {Q.~A.~T.}\ \bibnamefont {Nguyen}},
  \bibinfo {author} {\bibfnamefont {G.}~\bibnamefont {{Won Kim}}}, \bibinfo
  {author} {\bibfnamefont {M.}~\bibnamefont {{Hyeok Lee}}}, \bibinfo {author}
  {\bibfnamefont {S.}~\bibnamefont {{In Yoon}}}, \bibinfo {author}
  {\bibfnamefont {S.~H.}\ \bibnamefont {Rhim}},\ and\ \bibinfo {author}
  {\bibfnamefont {Y.}~\bibnamefont {{Keun Kim}}},\ }\bibfield  {title}
  {\bibinfo {title} {Enhanced spin–orbit torque efficiency with low
  resistivity in perpendicularly magnetized heterostructures consisting of
  si-alloyed \(\beta\)-w layers},\ }\href
  {https://doi.org/https://doi.org/10.1016/j.apsusc.2022.155352} {\bibfield
  {journal} {\bibinfo  {journal} {Applied Surface Science}\ }\textbf {\bibinfo
  {volume} {609}},\ \bibinfo {pages} {155352} (\bibinfo {year}
  {2023})}\BibitemShut {NoStop}%
\bibitem [{\citenamefont {Zhou}\ \emph {et~al.}(2025)\citenamefont {Zhou},
  \citenamefont {Poncé},\ and\ \citenamefont {Charlier}}]{Zhou_2025}%
  \BibitemOpen
  \bibfield  {author} {\bibinfo {author} {\bibfnamefont {J.}~\bibnamefont
  {Zhou}}, \bibinfo {author} {\bibfnamefont {S.}~\bibnamefont {Poncé}},\ and\
  \bibinfo {author} {\bibfnamefont {J.-C.}\ \bibnamefont {Charlier}},\
  }\bibfield  {title} {\bibinfo {title} {High-throughput calculations of spin
  hall conductivity in non-magnetic 2d materials},\ }\bibfield  {journal}
  {\bibinfo  {journal} {npj 2D Materials and Applications}\ }\textbf {\bibinfo
  {volume} {9}},\ \href {https://doi.org/10.1038/s41699-025-00562-4}
  {10.1038/s41699-025-00562-4} (\bibinfo {year} {2025})\BibitemShut {NoStop}%
\bibitem [{\citenamefont {Jungwirth}\ \emph {et~al.}(2012)\citenamefont
  {Jungwirth}, \citenamefont {Wunderlich},\ and\ \citenamefont
  {Olejn{\'\i}k}}]{jungwirth2012spin}%
  \BibitemOpen
  \bibfield  {author} {\bibinfo {author} {\bibfnamefont {T.}~\bibnamefont
  {Jungwirth}}, \bibinfo {author} {\bibfnamefont {J.}~\bibnamefont
  {Wunderlich}},\ and\ \bibinfo {author} {\bibfnamefont {K.}~\bibnamefont
  {Olejn{\'\i}k}},\ }\bibfield  {title} {\bibinfo {title} {Spin hall effect
  devices},\ }\href@noop {} {\bibfield  {journal} {\bibinfo  {journal} {Nature
  materials}\ }\textbf {\bibinfo {volume} {11}},\ \bibinfo {pages} {382}
  (\bibinfo {year} {2012})}\BibitemShut {NoStop}%
\bibitem [{\citenamefont {Hirsch}(1999)}]{PhysRevLett.83.1834}%
  \BibitemOpen
  \bibfield  {author} {\bibinfo {author} {\bibfnamefont {J.~E.}\ \bibnamefont
  {Hirsch}},\ }\bibfield  {title} {\bibinfo {title} {Spin hall effect},\ }\href
  {https://doi.org/10.1103/PhysRevLett.83.1834} {\bibfield  {journal} {\bibinfo
   {journal} {Phys. Rev. Lett.}\ }\textbf {\bibinfo {volume} {83}},\ \bibinfo
  {pages} {1834} (\bibinfo {year} {1999})}\BibitemShut {NoStop}%
\bibitem [{\citenamefont {Ferreira}\ \emph {et~al.}(2014)\citenamefont
  {Ferreira}, \citenamefont {Rappoport}, \citenamefont {Cazalilla},\ and\
  \citenamefont {Castro~Neto}}]{PhysRevLett.112.066601}%
  \BibitemOpen
  \bibfield  {author} {\bibinfo {author} {\bibfnamefont {A.}~\bibnamefont
  {Ferreira}}, \bibinfo {author} {\bibfnamefont {T.~G.}\ \bibnamefont
  {Rappoport}}, \bibinfo {author} {\bibfnamefont {M.~A.}\ \bibnamefont
  {Cazalilla}},\ and\ \bibinfo {author} {\bibfnamefont {A.~H.}\ \bibnamefont
  {Castro~Neto}},\ }\bibfield  {title} {\bibinfo {title} {Extrinsic spin hall
  effect induced by resonant skew scattering in graphene},\ }\href
  {https://doi.org/10.1103/PhysRevLett.112.066601} {\bibfield  {journal}
  {\bibinfo  {journal} {Phys. Rev. Lett.}\ }\textbf {\bibinfo {volume} {112}},\
  \bibinfo {pages} {066601} (\bibinfo {year} {2014})}\BibitemShut {NoStop}%
\bibitem [{\citenamefont {Zhang}(2000)}]{PhysRevLett.85.393}%
  \BibitemOpen
  \bibfield  {author} {\bibinfo {author} {\bibfnamefont {S.}~\bibnamefont
  {Zhang}},\ }\bibfield  {title} {\bibinfo {title} {Spin hall effect in the
  presence of spin diffusion},\ }\href
  {https://doi.org/10.1103/PhysRevLett.85.393} {\bibfield  {journal} {\bibinfo
  {journal} {Phys. Rev. Lett.}\ }\textbf {\bibinfo {volume} {85}},\ \bibinfo
  {pages} {393} (\bibinfo {year} {2000})}\BibitemShut {NoStop}%
\bibitem [{\citenamefont {Dyakonov}(1971)}]{dyakonov1971possibility}%
  \BibitemOpen
  \bibfield  {author} {\bibinfo {author} {\bibfnamefont {M.~I.}\ \bibnamefont
  {Dyakonov}},\ }\bibfield  {title} {\bibinfo {title} {Possibility of orienting
  electron spins with current},\ }\href@noop {} {\bibfield  {journal} {\bibinfo
   {journal} {JETP Lett. USSR}\ }\textbf {\bibinfo {volume} {13}},\ \bibinfo
  {pages} {467} (\bibinfo {year} {1971})}\BibitemShut {NoStop}%
\bibitem [{\citenamefont {Milletar\`{\i}}\ and\ \citenamefont
  {Ferreira}(2016)}]{PhysRevB.94.134202}%
  \BibitemOpen
  \bibfield  {author} {\bibinfo {author} {\bibfnamefont {M.}~\bibnamefont
  {Milletar\`{\i}}}\ and\ \bibinfo {author} {\bibfnamefont {A.}~\bibnamefont
  {Ferreira}},\ }\bibfield  {title} {\bibinfo {title} {Quantum diagrammatic
  theory of the extrinsic spin hall effect in graphene},\ }\href
  {https://doi.org/10.1103/PhysRevB.94.134202} {\bibfield  {journal} {\bibinfo
  {journal} {Phys. Rev. B}\ }\textbf {\bibinfo {volume} {94}},\ \bibinfo
  {pages} {134202} (\bibinfo {year} {2016})}\BibitemShut {NoStop}%
\bibitem [{\citenamefont {Sahoo}\ \emph {et~al.}(2025)\citenamefont {Sahoo},
  \citenamefont {Bhattacharjee},\ and\ \citenamefont
  {Muralidharan}}]{sahoo2025rashbainducedspinhallresponse}%
  \BibitemOpen
  \bibfield  {author} {\bibinfo {author} {\bibfnamefont {S.}~\bibnamefont
  {Sahoo}}, \bibinfo {author} {\bibfnamefont {S.}~\bibnamefont
  {Bhattacharjee}},\ and\ \bibinfo {author} {\bibfnamefont {B.}~\bibnamefont
  {Muralidharan}},\ }\href {https://arxiv.org/abs/2507.01584} {\bibinfo {title}
  {Rashba-induced spin hall response in a disordered $wte_2$ four-terminal
  structure}} (\bibinfo {year} {2025}),\ \Eprint
  {https://arxiv.org/abs/2507.01584} {arXiv:2507.01584 [cond-mat.mes-hall]}
  \BibitemShut {NoStop}%
\bibitem [{\citenamefont {Chi}\ \emph {et~al.}(2025)\citenamefont {Chi},
  \citenamefont {Dolan}, \citenamefont {Yang}, \citenamefont
  {Mart\'{\i}n-Garc\'{\i}a}, \citenamefont {Gobbi}, \citenamefont {Hueso},\
  and\ \citenamefont {Casanova}}]{k5tg-xm6q}%
  \BibitemOpen
  \bibfield  {author} {\bibinfo {author} {\bibfnamefont {Z.}~\bibnamefont
  {Chi}}, \bibinfo {author} {\bibfnamefont {E.}~\bibnamefont {Dolan}}, \bibinfo
  {author} {\bibfnamefont {H.}~\bibnamefont {Yang}}, \bibinfo {author}
  {\bibfnamefont {B.}~\bibnamefont {Mart\'{\i}n-Garc\'{\i}a}}, \bibinfo
  {author} {\bibfnamefont {M.}~\bibnamefont {Gobbi}}, \bibinfo {author}
  {\bibfnamefont {L.~E.}\ \bibnamefont {Hueso}},\ and\ \bibinfo {author}
  {\bibfnamefont {F.}~\bibnamefont {Casanova}},\ }\bibfield  {title} {\bibinfo
  {title} {Gate-tunable charge-spin interconversion in graphene/heavy-metal
  heterostructures},\ }\href {https://doi.org/10.1103/k5tg-xm6q} {\bibfield
  {journal} {\bibinfo  {journal} {Phys. Rev. Appl.}\ }\textbf {\bibinfo
  {volume} {24}},\ \bibinfo {pages} {064076} (\bibinfo {year}
  {2025})}\BibitemShut {NoStop}%
\bibitem [{\citenamefont {Ruan}\ \emph {et~al.}(2026)\citenamefont {Ruan},
  \citenamefont {Tao}, \citenamefont {Wu}, \citenamefont {Chen}, \citenamefont
  {Xu}, \citenamefont {Jiang},\ and\ \citenamefont {Meng}}]{c5x3-yqf8}%
  \BibitemOpen
  \bibfield  {author} {\bibinfo {author} {\bibfnamefont {Y.~Q.}\ \bibnamefont
  {Ruan}}, \bibinfo {author} {\bibfnamefont {S.~K.}\ \bibnamefont {Tao}},
  \bibinfo {author} {\bibfnamefont {Y.}~\bibnamefont {Wu}}, \bibinfo {author}
  {\bibfnamefont {J.~K.}\ \bibnamefont {Chen}}, \bibinfo {author}
  {\bibfnamefont {X.~G.}\ \bibnamefont {Xu}}, \bibinfo {author} {\bibfnamefont
  {Y.}~\bibnamefont {Jiang}},\ and\ \bibinfo {author} {\bibfnamefont {K.~K.}\
  \bibnamefont {Meng}},\ }\bibfield  {title} {\bibinfo {title} {Robust spin
  hall effect in polycrystalline pt-mn-sn thin films: Insights from experiments
  and kwant modeling},\ }\href {https://doi.org/10.1103/c5x3-yqf8} {\bibfield
  {journal} {\bibinfo  {journal} {Phys. Rev. Appl.}\ ,\ } (\bibinfo {year}
  {2026})}\BibitemShut {NoStop}%
\bibitem [{\citenamefont {Ingla-Ayn\'es}\ \emph {et~al.}(2015)\citenamefont
  {Ingla-Ayn\'es}, \citenamefont {Guimar\~aes}, \citenamefont {Meijerink},
  \citenamefont {Zomer},\ and\ \citenamefont {van Wees}}]{PhysRevB.92.201410}%
  \BibitemOpen
  \bibfield  {author} {\bibinfo {author} {\bibfnamefont {J.}~\bibnamefont
  {Ingla-Ayn\'es}}, \bibinfo {author} {\bibfnamefont {M.~H.~D.}\ \bibnamefont
  {Guimar\~aes}}, \bibinfo {author} {\bibfnamefont {R.~J.}\ \bibnamefont
  {Meijerink}}, \bibinfo {author} {\bibfnamefont {P.~J.}\ \bibnamefont
  {Zomer}},\ and\ \bibinfo {author} {\bibfnamefont {B.~J.}\ \bibnamefont {van
  Wees}},\ }\bibfield  {title} {\bibinfo {title}
  {$24\ensuremath{-}\ensuremath{\mu}\mathrm{m}$ spin relaxation length in boron
  nitride encapsulated bilayer graphene},\ }\href
  {https://doi.org/10.1103/PhysRevB.92.201410} {\bibfield  {journal} {\bibinfo
  {journal} {Phys. Rev. B}\ }\textbf {\bibinfo {volume} {92}},\ \bibinfo
  {pages} {201410} (\bibinfo {year} {2015})}\BibitemShut {NoStop}%
\bibitem [{\citenamefont {Dr{\"o}geler}\ \emph {et~al.}(2016)\citenamefont
  {Dr{\"o}geler}, \citenamefont {Franzen}, \citenamefont {Volmer},
  \citenamefont {Pohlmann}, \citenamefont {Banszerus}, \citenamefont {Wolter},
  \citenamefont {Watanabe}, \citenamefont {Taniguchi}, \citenamefont
  {Stampfer},\ and\ \citenamefont
  {Beschoten}}]{doi:10.1021/acs.nanolett.6b00497}%
  \BibitemOpen
  \bibfield  {author} {\bibinfo {author} {\bibfnamefont {M.}~\bibnamefont
  {Dr{\"o}geler}}, \bibinfo {author} {\bibfnamefont {C.}~\bibnamefont
  {Franzen}}, \bibinfo {author} {\bibfnamefont {F.}~\bibnamefont {Volmer}},
  \bibinfo {author} {\bibfnamefont {T.}~\bibnamefont {Pohlmann}}, \bibinfo
  {author} {\bibfnamefont {L.}~\bibnamefont {Banszerus}}, \bibinfo {author}
  {\bibfnamefont {M.}~\bibnamefont {Wolter}}, \bibinfo {author} {\bibfnamefont
  {K.}~\bibnamefont {Watanabe}}, \bibinfo {author} {\bibfnamefont
  {T.}~\bibnamefont {Taniguchi}}, \bibinfo {author} {\bibfnamefont
  {C.}~\bibnamefont {Stampfer}},\ and\ \bibinfo {author} {\bibfnamefont
  {B.}~\bibnamefont {Beschoten}},\ }\bibfield  {title} {\bibinfo {title} {Spin
  lifetimes exceeding 12 ns in graphene nonlocal spin valve devices},\ }\href
  {https://doi.org/10.1021/acs.nanolett.6b00497} {\bibfield  {journal}
  {\bibinfo  {journal} {Nano Letters}\ }\textbf {\bibinfo {volume} {16}},\
  \bibinfo {pages} {3533} (\bibinfo {year} {2016})},\ \bibinfo {note} {pMID:
  27210240},\ \Eprint
  {https://arxiv.org/abs/https://doi.org/10.1021/acs.nanolett.6b00497}
  {https://doi.org/10.1021/acs.nanolett.6b00497} \BibitemShut {NoStop}%
\bibitem [{\citenamefont {Yan}\ \emph {et~al.}(2016)\citenamefont {Yan},
  \citenamefont {Phillips}, \citenamefont {Barbone}, \citenamefont
  {H\"am\"al\"ainen}, \citenamefont {Lombardo}, \citenamefont {Ghidini},
  \citenamefont {Moya}, \citenamefont {Maccherozzi}, \citenamefont {van
  Dijken}, \citenamefont {Dhesi}, \citenamefont {Ferrari},\ and\ \citenamefont
  {Mathur}}]{PhysRevLett.117.147201}%
  \BibitemOpen
  \bibfield  {author} {\bibinfo {author} {\bibfnamefont {W.}~\bibnamefont
  {Yan}}, \bibinfo {author} {\bibfnamefont {L.~C.}\ \bibnamefont {Phillips}},
  \bibinfo {author} {\bibfnamefont {M.}~\bibnamefont {Barbone}}, \bibinfo
  {author} {\bibfnamefont {S.~J.}\ \bibnamefont {H\"am\"al\"ainen}}, \bibinfo
  {author} {\bibfnamefont {A.}~\bibnamefont {Lombardo}}, \bibinfo {author}
  {\bibfnamefont {M.}~\bibnamefont {Ghidini}}, \bibinfo {author} {\bibfnamefont
  {X.}~\bibnamefont {Moya}}, \bibinfo {author} {\bibfnamefont {F.}~\bibnamefont
  {Maccherozzi}}, \bibinfo {author} {\bibfnamefont {S.}~\bibnamefont {van
  Dijken}}, \bibinfo {author} {\bibfnamefont {S.~S.}\ \bibnamefont {Dhesi}},
  \bibinfo {author} {\bibfnamefont {A.~C.}\ \bibnamefont {Ferrari}},\ and\
  \bibinfo {author} {\bibfnamefont {N.~D.}\ \bibnamefont {Mathur}},\ }\bibfield
   {title} {\bibinfo {title} {Long spin diffusion length in few-layer graphene
  flakes},\ }\href {https://doi.org/10.1103/PhysRevLett.117.147201} {\bibfield
  {journal} {\bibinfo  {journal} {Phys. Rev. Lett.}\ }\textbf {\bibinfo
  {volume} {117}},\ \bibinfo {pages} {147201} (\bibinfo {year}
  {2016})}\BibitemShut {NoStop}%
\bibitem [{\citenamefont {Safeer}\ \emph {et~al.}(2019)\citenamefont {Safeer},
  \citenamefont {Ingla-Ayn{\'e}s}, \citenamefont {Herling}, \citenamefont
  {Garcia}, \citenamefont {Vila}, \citenamefont {Ontoso}, \citenamefont
  {Calvo}, \citenamefont {Roche}, \citenamefont {Hueso},\ and\ \citenamefont
  {Casanova}}]{doi:10.1021/acs.nanolett.8b04368}%
  \BibitemOpen
  \bibfield  {author} {\bibinfo {author} {\bibfnamefont {C.~K.}\ \bibnamefont
  {Safeer}}, \bibinfo {author} {\bibfnamefont {J.}~\bibnamefont
  {Ingla-Ayn{\'e}s}}, \bibinfo {author} {\bibfnamefont {F.}~\bibnamefont
  {Herling}}, \bibinfo {author} {\bibfnamefont {J.~H.}\ \bibnamefont {Garcia}},
  \bibinfo {author} {\bibfnamefont {M.}~\bibnamefont {Vila}}, \bibinfo {author}
  {\bibfnamefont {N.}~\bibnamefont {Ontoso}}, \bibinfo {author} {\bibfnamefont
  {M.~R.}\ \bibnamefont {Calvo}}, \bibinfo {author} {\bibfnamefont
  {S.}~\bibnamefont {Roche}}, \bibinfo {author} {\bibfnamefont {L.~E.}\
  \bibnamefont {Hueso}},\ and\ \bibinfo {author} {\bibfnamefont
  {F.}~\bibnamefont {Casanova}},\ }\bibfield  {title} {\bibinfo {title}
  {Room-temperature spin hall effect in graphene/mos2 van der waals
  heterostructures},\ }\href {https://doi.org/10.1021/acs.nanolett.8b04368}
  {\bibfield  {journal} {\bibinfo  {journal} {Nano Letters}\ }\textbf {\bibinfo
  {volume} {19}},\ \bibinfo {pages} {1074} (\bibinfo {year} {2019})},\ \Eprint
  {https://arxiv.org/abs/https://doi.org/10.1021/acs.nanolett.8b04368}
  {https://doi.org/10.1021/acs.nanolett.8b04368} \BibitemShut {NoStop}%
\bibitem [{\citenamefont {Ingla-Ayn\'es}\ \emph {et~al.}(2021)\citenamefont
  {Ingla-Ayn\'es}, \citenamefont {Herling}, \citenamefont {Fabian},
  \citenamefont {Hueso},\ and\ \citenamefont
  {Casanova}}]{PhysRevLett.127.047202}%
  \BibitemOpen
  \bibfield  {author} {\bibinfo {author} {\bibfnamefont {J.}~\bibnamefont
  {Ingla-Ayn\'es}}, \bibinfo {author} {\bibfnamefont {F.}~\bibnamefont
  {Herling}}, \bibinfo {author} {\bibfnamefont {J.}~\bibnamefont {Fabian}},
  \bibinfo {author} {\bibfnamefont {L.~E.}\ \bibnamefont {Hueso}},\ and\
  \bibinfo {author} {\bibfnamefont {F.}~\bibnamefont {Casanova}},\ }\bibfield
  {title} {\bibinfo {title} {Electrical control of valley-zeeman
  spin-orbit-coupling--induced spin precession at room temperature},\ }\href
  {https://doi.org/10.1103/PhysRevLett.127.047202} {\bibfield  {journal}
  {\bibinfo  {journal} {Phys. Rev. Lett.}\ }\textbf {\bibinfo {volume} {127}},\
  \bibinfo {pages} {047202} (\bibinfo {year} {2021})}\BibitemShut {NoStop}%
\bibitem [{\citenamefont {Tahir}\ \emph {et~al.}(2025)\citenamefont {Tahir},
  \citenamefont {Das}, \citenamefont {Gupta}, \citenamefont {Medwal},\ and\
  \citenamefont {Mukhopadhyay}}]{doi:10.1021/acsaelm.5c00754}%
  \BibitemOpen
  \bibfield  {author} {\bibinfo {author} {\bibfnamefont {M.}~\bibnamefont
  {Tahir}}, \bibinfo {author} {\bibfnamefont {S.}~\bibnamefont {Das}}, \bibinfo
  {author} {\bibfnamefont {M.}~\bibnamefont {Gupta}}, \bibinfo {author}
  {\bibfnamefont {R.}~\bibnamefont {Medwal}},\ and\ \bibinfo {author}
  {\bibfnamefont {S.}~\bibnamefont {Mukhopadhyay}},\ }\bibfield  {title}
  {\bibinfo {title} {Harnessing the graphene/pt/fm interface for improved
  spin-to-charge current conversion},\ }\href
  {https://doi.org/10.1021/acsaelm.5c00754} {\bibfield  {journal} {\bibinfo
  {journal} {ACS Applied Electronic Materials}\ }\textbf {\bibinfo {volume}
  {7}},\ \bibinfo {pages} {6450} (\bibinfo {year} {2025})},\ \Eprint
  {https://arxiv.org/abs/https://doi.org/10.1021/acsaelm.5c00754}
  {https://doi.org/10.1021/acsaelm.5c00754} \BibitemShut {NoStop}%
\bibitem [{\citenamefont {Ramos}\ \emph {et~al.}(2018)\citenamefont {Ramos},
  \citenamefont {Vasconcelos},\ and\ \citenamefont {Barbosa}}]{ramo}%
  \BibitemOpen
  \bibfield  {author} {\bibinfo {author} {\bibfnamefont {J.}~\bibnamefont
  {Ramos}}, \bibinfo {author} {\bibfnamefont {T.}~\bibnamefont {Vasconcelos}},\
  and\ \bibinfo {author} {\bibfnamefont {A.}~\bibnamefont {Barbosa}},\
  }\bibfield  {title} {\bibinfo {title} {Spin-to-charge conversion in 2d
  electron gas and single-layer graphene devices},\ }\href@noop {} {\bibfield
  {journal} {\bibinfo  {journal} {Journal of Applied Physics}\ }\textbf
  {\bibinfo {volume} {123}},\ \bibinfo {pages} {034304} (\bibinfo {year}
  {2018})}\BibitemShut {NoStop}%
\bibitem [{\citenamefont {Mendes}\ \emph {et~al.}(2019)\citenamefont {Mendes},
  \citenamefont {Alves~Santos}, \citenamefont {Chagas}, \citenamefont
  {Magalhães-Paniago}, \citenamefont {Mori}, \citenamefont {Holanda},
  \citenamefont {Meireles}, \citenamefont {Lacerda}, \citenamefont {Azevedo},\
  and\ \citenamefont {Rezende}}]{Mendes_2019}%
  \BibitemOpen
  \bibfield  {author} {\bibinfo {author} {\bibfnamefont {J.~B.~S.}\
  \bibnamefont {Mendes}}, \bibinfo {author} {\bibfnamefont {O.}~\bibnamefont
  {Alves~Santos}}, \bibinfo {author} {\bibfnamefont {T.}~\bibnamefont
  {Chagas}}, \bibinfo {author} {\bibfnamefont {R.}~\bibnamefont
  {Magalhães-Paniago}}, \bibinfo {author} {\bibfnamefont {T.~J.~A.}\
  \bibnamefont {Mori}}, \bibinfo {author} {\bibfnamefont {J.}~\bibnamefont
  {Holanda}}, \bibinfo {author} {\bibfnamefont {L.~M.}\ \bibnamefont
  {Meireles}}, \bibinfo {author} {\bibfnamefont {R.~G.}\ \bibnamefont
  {Lacerda}}, \bibinfo {author} {\bibfnamefont {A.}~\bibnamefont {Azevedo}},\
  and\ \bibinfo {author} {\bibfnamefont {S.~M.}\ \bibnamefont {Rezende}},\
  }\bibfield  {title} {\bibinfo {title} {Direct detection of induced magnetic
  moment and efficient spin-to-charge conversion in graphene/ferromagnetic
  structures},\ }\bibfield  {journal} {\bibinfo  {journal} {Physical Review B}\
  }\textbf {\bibinfo {volume} {99}},\ \href
  {https://doi.org/10.1103/physrevb.99.214446} {10.1103/physrevb.99.214446}
  (\bibinfo {year} {2019})\BibitemShut {NoStop}%
\bibitem [{\citenamefont {Dueñas}\ \emph {et~al.}(2025)\citenamefont
  {Dueñas}, \citenamefont {de~Castro}, \citenamefont {Garcia},\ and\
  \citenamefont {Roche}}]{dueas2025optimalspinchargeinterconversiongraphene}%
  \BibitemOpen
  \bibfield  {author} {\bibinfo {author} {\bibfnamefont {J.~M.}\ \bibnamefont
  {Dueñas}}, \bibinfo {author} {\bibfnamefont {S.~G.}\ \bibnamefont
  {de~Castro}}, \bibinfo {author} {\bibfnamefont {J.~H.}\ \bibnamefont
  {Garcia}},\ and\ \bibinfo {author} {\bibfnamefont {S.}~\bibnamefont
  {Roche}},\ }\bibfield  {title} {\bibinfo {title} {Optimal spin-charge
  interconversion in graphene through spin-pseudospin entanglement control},\
  }\bibfield  {journal} {\bibinfo  {journal} {Communications Physics}\ }\href
  {https://doi.org/10.1038/s42005-026-02658-9} {10.1038/s42005-026-02658-9}
  (\bibinfo {year} {2025})\BibitemShut {NoStop}%
\bibitem [{\citenamefont {Gmitra}\ \emph {et~al.}(2013)\citenamefont {Gmitra},
  \citenamefont {Kochan},\ and\ \citenamefont
  {Fabian}}]{PhysRevLett.110.246602}%
  \BibitemOpen
  \bibfield  {author} {\bibinfo {author} {\bibfnamefont {M.}~\bibnamefont
  {Gmitra}}, \bibinfo {author} {\bibfnamefont {D.}~\bibnamefont {Kochan}},\
  and\ \bibinfo {author} {\bibfnamefont {J.}~\bibnamefont {Fabian}},\
  }\bibfield  {title} {\bibinfo {title} {Spin-orbit coupling in hydrogenated
  graphene},\ }\href {https://doi.org/10.1103/PhysRevLett.110.246602}
  {\bibfield  {journal} {\bibinfo  {journal} {Phys. Rev. Lett.}\ }\textbf
  {\bibinfo {volume} {110}},\ \bibinfo {pages} {246602} (\bibinfo {year}
  {2013})}\BibitemShut {NoStop}%
\bibitem [{\citenamefont {Castro~Neto}\ and\ \citenamefont
  {Guinea}(2009)}]{PhysRevLett.103.026804}%
  \BibitemOpen
  \bibfield  {author} {\bibinfo {author} {\bibfnamefont {A.~H.}\ \bibnamefont
  {Castro~Neto}}\ and\ \bibinfo {author} {\bibfnamefont {F.}~\bibnamefont
  {Guinea}},\ }\bibfield  {title} {\bibinfo {title} {Impurity-induced
  spin-orbit coupling in graphene},\ }\href
  {https://doi.org/10.1103/PhysRevLett.103.026804} {\bibfield  {journal}
  {\bibinfo  {journal} {Phys. Rev. Lett.}\ }\textbf {\bibinfo {volume} {103}},\
  \bibinfo {pages} {026804} (\bibinfo {year} {2009})}\BibitemShut {NoStop}%
\bibitem [{\citenamefont {Gmitra}\ and\ \citenamefont
  {Fabian}(2015)}]{PhysRevB.92.155403}%
  \BibitemOpen
  \bibfield  {author} {\bibinfo {author} {\bibfnamefont {M.}~\bibnamefont
  {Gmitra}}\ and\ \bibinfo {author} {\bibfnamefont {J.}~\bibnamefont
  {Fabian}},\ }\bibfield  {title} {\bibinfo {title} {Graphene on
  transition-metal dichalcogenides: A platform for proximity spin-orbit physics
  and optospintronics},\ }\href {https://doi.org/10.1103/PhysRevB.92.155403}
  {\bibfield  {journal} {\bibinfo  {journal} {Phys. Rev. B}\ }\textbf {\bibinfo
  {volume} {92}},\ \bibinfo {pages} {155403} (\bibinfo {year}
  {2015})}\BibitemShut {NoStop}%
\bibitem [{\citenamefont {Gmitra}\ \emph {et~al.}(2016)\citenamefont {Gmitra},
  \citenamefont {Kochan}, \citenamefont {H\"ogl},\ and\ \citenamefont
  {Fabian}}]{PhysRevB.93.155104}%
  \BibitemOpen
  \bibfield  {author} {\bibinfo {author} {\bibfnamefont {M.}~\bibnamefont
  {Gmitra}}, \bibinfo {author} {\bibfnamefont {D.}~\bibnamefont {Kochan}},
  \bibinfo {author} {\bibfnamefont {P.}~\bibnamefont {H\"ogl}},\ and\ \bibinfo
  {author} {\bibfnamefont {J.}~\bibnamefont {Fabian}},\ }\bibfield  {title}
  {\bibinfo {title} {Trivial and inverted dirac bands and the emergence of
  quantum spin hall states in graphene on transition-metal dichalcogenides},\
  }\href {https://doi.org/10.1103/PhysRevB.93.155104} {\bibfield  {journal}
  {\bibinfo  {journal} {Phys. Rev. B}\ }\textbf {\bibinfo {volume} {93}},\
  \bibinfo {pages} {155104} (\bibinfo {year} {2016})}\BibitemShut {NoStop}%
\bibitem [{\citenamefont {Gmitra}\ and\ \citenamefont
  {Fabian}(2017)}]{PhysRevLett.119.146401}%
  \BibitemOpen
  \bibfield  {author} {\bibinfo {author} {\bibfnamefont {M.}~\bibnamefont
  {Gmitra}}\ and\ \bibinfo {author} {\bibfnamefont {J.}~\bibnamefont
  {Fabian}},\ }\bibfield  {title} {\bibinfo {title} {Proximity effects in
  bilayer graphene on monolayer ${\mathrm{wse}}_{2}$: Field-effect spin valley
  locking, spin-orbit valve, and spin transistor},\ }\href
  {https://doi.org/10.1103/PhysRevLett.119.146401} {\bibfield  {journal}
  {\bibinfo  {journal} {Phys. Rev. Lett.}\ }\textbf {\bibinfo {volume} {119}},\
  \bibinfo {pages} {146401} (\bibinfo {year} {2017})}\BibitemShut {NoStop}%
\bibitem [{\citenamefont {Garcia}\ \emph {et~al.}(2018)\citenamefont {Garcia},
  \citenamefont {Vila}, \citenamefont {Cummings},\ and\ \citenamefont
  {Roche}}]{garcia2018spin}%
  \BibitemOpen
  \bibfield  {author} {\bibinfo {author} {\bibfnamefont {J.~H.}\ \bibnamefont
  {Garcia}}, \bibinfo {author} {\bibfnamefont {M.}~\bibnamefont {Vila}},
  \bibinfo {author} {\bibfnamefont {A.~W.}\ \bibnamefont {Cummings}},\ and\
  \bibinfo {author} {\bibfnamefont {S.}~\bibnamefont {Roche}},\ }\bibfield
  {title} {\bibinfo {title} {Spin transport in graphene/transition metal
  dichalcogenide heterostructures},\ }\href@noop {} {\bibfield  {journal}
  {\bibinfo  {journal} {Chemical Society Reviews}\ }\textbf {\bibinfo {volume}
  {47}},\ \bibinfo {pages} {3359} (\bibinfo {year} {2018})}\BibitemShut
  {NoStop}%
\bibitem [{\citenamefont {Yang}\ \emph
  {et~al.}(2024{\natexlab{a}})\citenamefont {Yang}, \citenamefont
  {Martín-García}, \citenamefont {Kimák}, \citenamefont {Schmoranzerová},
  \citenamefont {Dolan}, \citenamefont {Chi}, \citenamefont {Gobbi},
  \citenamefont {Němec}, \citenamefont {Hueso},\ and\ \citenamefont
  {Casanova}}]{Yang_2024}%
  \BibitemOpen
  \bibfield  {author} {\bibinfo {author} {\bibfnamefont {H.}~\bibnamefont
  {Yang}}, \bibinfo {author} {\bibfnamefont {B.}~\bibnamefont
  {Martín-García}}, \bibinfo {author} {\bibfnamefont {J.}~\bibnamefont
  {Kimák}}, \bibinfo {author} {\bibfnamefont {E.}~\bibnamefont
  {Schmoranzerová}}, \bibinfo {author} {\bibfnamefont {E.}~\bibnamefont
  {Dolan}}, \bibinfo {author} {\bibfnamefont {Z.}~\bibnamefont {Chi}}, \bibinfo
  {author} {\bibfnamefont {M.}~\bibnamefont {Gobbi}}, \bibinfo {author}
  {\bibfnamefont {P.}~\bibnamefont {Němec}}, \bibinfo {author} {\bibfnamefont
  {L.~E.}\ \bibnamefont {Hueso}},\ and\ \bibinfo {author} {\bibfnamefont
  {F.}~\bibnamefont {Casanova}},\ }\bibfield  {title} {\bibinfo {title}
  {Twist-angle-tunable spin texture in wse2/graphene van der waals
  heterostructures},\ }\href {https://doi.org/10.1038/s41563-024-01985-y}
  {\bibfield  {journal} {\bibinfo  {journal} {Nature Materials}\ }\textbf
  {\bibinfo {volume} {23}},\ \bibinfo {pages} {1502–1508} (\bibinfo {year}
  {2024}{\natexlab{a}})}\BibitemShut {NoStop}%
\bibitem [{\citenamefont {Yang}\ \emph
  {et~al.}(2024{\natexlab{b}})\citenamefont {Yang}, \citenamefont {Chi},
  \citenamefont {Avedissian}, \citenamefont {Dolan}, \citenamefont
  {Karuppasamy}, \citenamefont {Martín-García}, \citenamefont {Gobbi},
  \citenamefont {Sofer}, \citenamefont {Hueso},\ and\ \citenamefont
  {Casanova}}]{https://doi.org/10.1002/adfm.202404872}%
  \BibitemOpen
  \bibfield  {author} {\bibinfo {author} {\bibfnamefont {H.}~\bibnamefont
  {Yang}}, \bibinfo {author} {\bibfnamefont {Z.}~\bibnamefont {Chi}}, \bibinfo
  {author} {\bibfnamefont {G.}~\bibnamefont {Avedissian}}, \bibinfo {author}
  {\bibfnamefont {E.}~\bibnamefont {Dolan}}, \bibinfo {author} {\bibfnamefont
  {M.}~\bibnamefont {Karuppasamy}}, \bibinfo {author} {\bibfnamefont
  {B.}~\bibnamefont {Martín-García}}, \bibinfo {author} {\bibfnamefont
  {M.}~\bibnamefont {Gobbi}}, \bibinfo {author} {\bibfnamefont
  {Z.}~\bibnamefont {Sofer}}, \bibinfo {author} {\bibfnamefont {L.~E.}\
  \bibnamefont {Hueso}},\ and\ \bibinfo {author} {\bibfnamefont
  {F.}~\bibnamefont {Casanova}},\ }\bibfield  {title} {\bibinfo {title}
  {Gate-tunable spin hall effect in trilayer graphene/group-iv monochalcogenide
  van der waals heterostructures},\ }\href
  {https://doi.org/https://doi.org/10.1002/adfm.202404872} {\bibfield
  {journal} {\bibinfo  {journal} {Advanced Functional Materials}\ }\textbf
  {\bibinfo {volume} {34}},\ \bibinfo {pages} {2404872} (\bibinfo {year}
  {2024}{\natexlab{b}})},\ \Eprint
  {https://arxiv.org/abs/https://advanced.onlinelibrary.wiley.com/doi/pdf/10.1002/adfm.202404872}
  {https://advanced.onlinelibrary.wiley.com/doi/pdf/10.1002/adfm.202404872}
  \BibitemShut {NoStop}%
\bibitem [{\citenamefont {Ontoso}\ \emph {et~al.}(2023)\citenamefont {Ontoso},
  \citenamefont {Safeer}, \citenamefont {Herling}, \citenamefont
  {Ingla-Ayn\'es}, \citenamefont {Yang}, \citenamefont {Chi}, \citenamefont
  {Martin-Garcia}, \citenamefont {Robredo}, \citenamefont {Vergniory},
  \citenamefont {de~Juan}, \citenamefont {Reyes~Calvo}, \citenamefont {Hueso},\
  and\ \citenamefont {Casanova}}]{PhysRevApplied.19.014053}%
  \BibitemOpen
  \bibfield  {author} {\bibinfo {author} {\bibfnamefont {N.}~\bibnamefont
  {Ontoso}}, \bibinfo {author} {\bibfnamefont {C.~K.}\ \bibnamefont {Safeer}},
  \bibinfo {author} {\bibfnamefont {F.}~\bibnamefont {Herling}}, \bibinfo
  {author} {\bibfnamefont {J.}~\bibnamefont {Ingla-Ayn\'es}}, \bibinfo {author}
  {\bibfnamefont {H.}~\bibnamefont {Yang}}, \bibinfo {author} {\bibfnamefont
  {Z.}~\bibnamefont {Chi}}, \bibinfo {author} {\bibfnamefont {B.}~\bibnamefont
  {Martin-Garcia}}, \bibinfo {author} {\bibfnamefont {I.~n.}\ \bibnamefont
  {Robredo}}, \bibinfo {author} {\bibfnamefont {M.~G.}\ \bibnamefont
  {Vergniory}}, \bibinfo {author} {\bibfnamefont {F.}~\bibnamefont {de~Juan}},
  \bibinfo {author} {\bibfnamefont {M.}~\bibnamefont {Reyes~Calvo}}, \bibinfo
  {author} {\bibfnamefont {L.~E.}\ \bibnamefont {Hueso}},\ and\ \bibinfo
  {author} {\bibfnamefont {F.}~\bibnamefont {Casanova}},\ }\bibfield  {title}
  {\bibinfo {title} {Unconventional charge-to-spin conversion in
  graphene/${\mathrm{mo}\mathrm{te}}_{2}$ van der waals heterostructures},\
  }\href {https://doi.org/10.1103/PhysRevApplied.19.014053} {\bibfield
  {journal} {\bibinfo  {journal} {Phys. Rev. Appl.}\ }\textbf {\bibinfo
  {volume} {19}},\ \bibinfo {pages} {014053} (\bibinfo {year}
  {2023})}\BibitemShut {NoStop}%
\bibitem [{\citenamefont {Perkins}\ \emph {et~al.}(2024)\citenamefont
  {Perkins}, \citenamefont {Veneri},\ and\ \citenamefont
  {Ferreira}}]{PhysRevB.109.L241404}%
  \BibitemOpen
  \bibfield  {author} {\bibinfo {author} {\bibfnamefont {D.~T.~S.}\
  \bibnamefont {Perkins}}, \bibinfo {author} {\bibfnamefont {A.}~\bibnamefont
  {Veneri}},\ and\ \bibinfo {author} {\bibfnamefont {A.}~\bibnamefont
  {Ferreira}},\ }\bibfield  {title} {\bibinfo {title} {Spin hall effect:
  Symmetry breaking, twisting, and giant disorder renormalization},\ }\href
  {https://doi.org/10.1103/PhysRevB.109.L241404} {\bibfield  {journal}
  {\bibinfo  {journal} {Phys. Rev. B}\ }\textbf {\bibinfo {volume} {109}},\
  \bibinfo {pages} {L241404} (\bibinfo {year} {2024})}\BibitemShut {NoStop}%
\bibitem [{\citenamefont {Monaco}\ \emph {et~al.}(2021)\citenamefont {Monaco},
  \citenamefont {Ferreira},\ and\ \citenamefont
  {Raimondi}}]{PhysRevResearch.3.033137}%
  \BibitemOpen
  \bibfield  {author} {\bibinfo {author} {\bibfnamefont {C.}~\bibnamefont
  {Monaco}}, \bibinfo {author} {\bibfnamefont {A.}~\bibnamefont {Ferreira}},\
  and\ \bibinfo {author} {\bibfnamefont {R.}~\bibnamefont {Raimondi}},\
  }\bibfield  {title} {\bibinfo {title} {Spin hall and inverse spin galvanic
  effects in graphene with strong interfacial spin-orbit coupling: A
  quasi-classical green's function approach},\ }\href
  {https://doi.org/10.1103/PhysRevResearch.3.033137} {\bibfield  {journal}
  {\bibinfo  {journal} {Phys. Rev. Res.}\ }\textbf {\bibinfo {volume} {3}},\
  \bibinfo {pages} {033137} (\bibinfo {year} {2021})}\BibitemShut {NoStop}%
\bibitem [{\citenamefont {Barbosa}\ \emph {et~al.}(2021)\citenamefont
  {Barbosa}, \citenamefont {Ramos},\ and\ \citenamefont
  {Ferreira}}]{PhysRevB.103.L081111}%
  \BibitemOpen
  \bibfield  {author} {\bibinfo {author} {\bibfnamefont {A.~L.~R.}\
  \bibnamefont {Barbosa}}, \bibinfo {author} {\bibfnamefont {J.~G. G.~S.}\
  \bibnamefont {Ramos}},\ and\ \bibinfo {author} {\bibfnamefont
  {A.}~\bibnamefont {Ferreira}},\ }\bibfield  {title} {\bibinfo {title} {Effect
  of proximity-induced spin-orbit coupling in graphene mesoscopic billiards},\
  }\href {https://doi.org/10.1103/PhysRevB.103.L081111} {\bibfield  {journal}
  {\bibinfo  {journal} {Phys. Rev. B}\ }\textbf {\bibinfo {volume} {103}},\
  \bibinfo {pages} {L081111} (\bibinfo {year} {2021})}\BibitemShut {NoStop}%
\bibitem [{\citenamefont {Cunha}\ \emph {et~al.}(2024)\citenamefont {Cunha},
  \citenamefont {Garcia-Basabe}, \citenamefont {Larrude}, \citenamefont
  {Gamino}, \citenamefont {N.~Lima}, \citenamefont {Crasto~de Lima},
  \citenamefont {Fazzio}, \citenamefont {Rezende}, \citenamefont {Azevedo},\
  and\ \citenamefont {Mendes}}]{doi:10.1021/acsami.4c08539}%
  \BibitemOpen
  \bibfield  {author} {\bibinfo {author} {\bibfnamefont {R.~O.}\ \bibnamefont
  {Cunha}}, \bibinfo {author} {\bibfnamefont {Y.}~\bibnamefont
  {Garcia-Basabe}}, \bibinfo {author} {\bibfnamefont {D.~G.}\ \bibnamefont
  {Larrude}}, \bibinfo {author} {\bibfnamefont {M.}~\bibnamefont {Gamino}},
  \bibinfo {author} {\bibfnamefont {E.}~\bibnamefont {N.~Lima}}, \bibinfo
  {author} {\bibfnamefont {F.}~\bibnamefont {Crasto~de Lima}}, \bibinfo
  {author} {\bibfnamefont {A.}~\bibnamefont {Fazzio}}, \bibinfo {author}
  {\bibfnamefont {S.~M.}\ \bibnamefont {Rezende}}, \bibinfo {author}
  {\bibfnamefont {A.}~\bibnamefont {Azevedo}},\ and\ \bibinfo {author}
  {\bibfnamefont {J.~B.~S.}\ \bibnamefont {Mendes}},\ }\bibfield  {title}
  {\bibinfo {title} {Unraveling the spin-to-charge current conversion mechanism
  and charge transfer dynamics at the interface of graphene/ws2
  heterostructures at room temperature},\ }\href
  {https://doi.org/10.1021/acsami.4c08539} {\bibfield  {journal} {\bibinfo
  {journal} {ACS Applied Materials \& Interfaces}\ }\textbf {\bibinfo {volume}
  {16}},\ \bibinfo {pages} {56211} (\bibinfo {year} {2024})},\ \bibinfo {note}
  {pMID: 39356804},\ \Eprint
  {https://arxiv.org/abs/https://doi.org/10.1021/acsami.4c08539}
  {https://doi.org/10.1021/acsami.4c08539} \BibitemShut {NoStop}%
\bibitem [{\citenamefont {Ghiasi}\ \emph {et~al.}(2019)\citenamefont {Ghiasi},
  \citenamefont {Kaverzin}, \citenamefont {Blah},\ and\ \citenamefont {van
  Wees}}]{doi:10.1021/acs.nanolett.9b01611}%
  \BibitemOpen
  \bibfield  {author} {\bibinfo {author} {\bibfnamefont {T.~S.}\ \bibnamefont
  {Ghiasi}}, \bibinfo {author} {\bibfnamefont {A.~A.}\ \bibnamefont
  {Kaverzin}}, \bibinfo {author} {\bibfnamefont {P.~J.}\ \bibnamefont {Blah}},\
  and\ \bibinfo {author} {\bibfnamefont {B.~J.}\ \bibnamefont {van Wees}},\
  }\bibfield  {title} {\bibinfo {title} {Charge-to-spin conversion by the
  rashba–edelstein effect in two-dimensional van der waals heterostructures
  up to room temperature},\ }\href
  {https://doi.org/10.1021/acs.nanolett.9b01611} {\bibfield  {journal}
  {\bibinfo  {journal} {Nano Letters}\ }\textbf {\bibinfo {volume} {19}},\
  \bibinfo {pages} {5959} (\bibinfo {year} {2019})},\ \bibinfo {note} {pMID:
  31408607},\ \Eprint
  {https://arxiv.org/abs/https://doi.org/10.1021/acs.nanolett.9b01611}
  {https://doi.org/10.1021/acs.nanolett.9b01611} \BibitemShut {NoStop}%
\bibitem [{\citenamefont {Offidani}\ \emph {et~al.}(2017)\citenamefont
  {Offidani}, \citenamefont {Milletar\`{\i}}, \citenamefont {Raimondi},\ and\
  \citenamefont {Ferreira}}]{PhysRevLett.119.196801}%
  \BibitemOpen
  \bibfield  {author} {\bibinfo {author} {\bibfnamefont {M.}~\bibnamefont
  {Offidani}}, \bibinfo {author} {\bibfnamefont {M.}~\bibnamefont
  {Milletar\`{\i}}}, \bibinfo {author} {\bibfnamefont {R.}~\bibnamefont
  {Raimondi}},\ and\ \bibinfo {author} {\bibfnamefont {A.}~\bibnamefont
  {Ferreira}},\ }\bibfield  {title} {\bibinfo {title} {Optimal charge-to-spin
  conversion in graphene on transition-metal dichalcogenides},\ }\href
  {https://doi.org/10.1103/PhysRevLett.119.196801} {\bibfield  {journal}
  {\bibinfo  {journal} {Phys. Rev. Lett.}\ }\textbf {\bibinfo {volume} {119}},\
  \bibinfo {pages} {196801} (\bibinfo {year} {2017})}\BibitemShut {NoStop}%
\bibitem [{\citenamefont {Benítez}\ \emph {et~al.}(2020)\citenamefont
  {Benítez}, \citenamefont {Savero~Torres}, \citenamefont {Sierra},
  \citenamefont {Timmermans}, \citenamefont {Garcia}, \citenamefont {Roche},
  \citenamefont {Costache},\ and\ \citenamefont {Valenzuela}}]{Ben_tez_2020}%
  \BibitemOpen
  \bibfield  {author} {\bibinfo {author} {\bibfnamefont {L.~A.}\ \bibnamefont
  {Benítez}}, \bibinfo {author} {\bibfnamefont {W.}~\bibnamefont
  {Savero~Torres}}, \bibinfo {author} {\bibfnamefont {J.~F.}\ \bibnamefont
  {Sierra}}, \bibinfo {author} {\bibfnamefont {M.}~\bibnamefont {Timmermans}},
  \bibinfo {author} {\bibfnamefont {J.~H.}\ \bibnamefont {Garcia}}, \bibinfo
  {author} {\bibfnamefont {S.}~\bibnamefont {Roche}}, \bibinfo {author}
  {\bibfnamefont {M.~V.}\ \bibnamefont {Costache}},\ and\ \bibinfo {author}
  {\bibfnamefont {S.~O.}\ \bibnamefont {Valenzuela}},\ }\bibfield  {title}
  {\bibinfo {title} {Tunable room-temperature spin galvanic and spin hall
  effects in van der waals heterostructures},\ }\href
  {https://doi.org/10.1038/s41563-019-0575-1} {\bibfield  {journal} {\bibinfo
  {journal} {Nature Materials}\ }\textbf {\bibinfo {volume} {19}},\ \bibinfo
  {pages} {170–175} (\bibinfo {year} {2020})}\BibitemShut {NoStop}%
\bibitem [{\citenamefont {Li}\ \emph {et~al.}(2020)\citenamefont {Li},
  \citenamefont {Zhang}, \citenamefont {Myeong}, \citenamefont {Shin},
  \citenamefont {Lim}, \citenamefont {Kim}, \citenamefont {Kim}, \citenamefont
  {Jin}, \citenamefont {Cavill}, \citenamefont {Kim}, \citenamefont {Kim},
  \citenamefont {Lischner}, \citenamefont {Ferreira},\ and\ \citenamefont
  {Cho}}]{doi:10.1021/acsnano.0c01037}%
  \BibitemOpen
  \bibfield  {author} {\bibinfo {author} {\bibfnamefont {L.}~\bibnamefont
  {Li}}, \bibinfo {author} {\bibfnamefont {J.}~\bibnamefont {Zhang}}, \bibinfo
  {author} {\bibfnamefont {G.}~\bibnamefont {Myeong}}, \bibinfo {author}
  {\bibfnamefont {W.}~\bibnamefont {Shin}}, \bibinfo {author} {\bibfnamefont
  {H.}~\bibnamefont {Lim}}, \bibinfo {author} {\bibfnamefont {B.}~\bibnamefont
  {Kim}}, \bibinfo {author} {\bibfnamefont {S.}~\bibnamefont {Kim}}, \bibinfo
  {author} {\bibfnamefont {T.}~\bibnamefont {Jin}}, \bibinfo {author}
  {\bibfnamefont {S.}~\bibnamefont {Cavill}}, \bibinfo {author} {\bibfnamefont
  {B.~S.}\ \bibnamefont {Kim}}, \bibinfo {author} {\bibfnamefont
  {C.}~\bibnamefont {Kim}}, \bibinfo {author} {\bibfnamefont {J.}~\bibnamefont
  {Lischner}}, \bibinfo {author} {\bibfnamefont {A.}~\bibnamefont {Ferreira}},\
  and\ \bibinfo {author} {\bibfnamefont {S.}~\bibnamefont {Cho}},\ }\bibfield
  {title} {\bibinfo {title} {Gate-tunable reversible rashba–edelstein effect
  in a few-layer graphene/2h-tas2 heterostructure at room temperature},\ }\href
  {https://doi.org/10.1021/acsnano.0c01037} {\bibfield  {journal} {\bibinfo
  {journal} {ACS Nano}\ }\textbf {\bibinfo {volume} {14}},\ \bibinfo {pages}
  {5251} (\bibinfo {year} {2020})},\ \bibinfo {note} {pMID: 32267673},\ \Eprint
  {https://arxiv.org/abs/https://doi.org/10.1021/acsnano.0c01037}
  {https://doi.org/10.1021/acsnano.0c01037} \BibitemShut {NoStop}%
\bibitem [{\citenamefont {Cavill}\ \emph {et~al.}(2020)\citenamefont {Cavill},
  \citenamefont {Huang}, \citenamefont {Offidani}, \citenamefont {Lin},
  \citenamefont {Cazalilla},\ and\ \citenamefont
  {Ferreira}}]{PhysRevLett.124.236803}%
  \BibitemOpen
  \bibfield  {author} {\bibinfo {author} {\bibfnamefont {S.~A.}\ \bibnamefont
  {Cavill}}, \bibinfo {author} {\bibfnamefont {C.}~\bibnamefont {Huang}},
  \bibinfo {author} {\bibfnamefont {M.}~\bibnamefont {Offidani}}, \bibinfo
  {author} {\bibfnamefont {Y.-H.}\ \bibnamefont {Lin}}, \bibinfo {author}
  {\bibfnamefont {M.~A.}\ \bibnamefont {Cazalilla}},\ and\ \bibinfo {author}
  {\bibfnamefont {A.}~\bibnamefont {Ferreira}},\ }\bibfield  {title} {\bibinfo
  {title} {Proposal for unambiguous electrical detection of spin-charge
  conversion in lateral spin valves},\ }\href
  {https://doi.org/10.1103/PhysRevLett.124.236803} {\bibfield  {journal}
  {\bibinfo  {journal} {Phys. Rev. Lett.}\ }\textbf {\bibinfo {volume} {124}},\
  \bibinfo {pages} {236803} (\bibinfo {year} {2020})}\BibitemShut {NoStop}%
\bibitem [{\citenamefont {Khokhriakov}\ \emph {et~al.}(2020)\citenamefont
  {Khokhriakov}, \citenamefont {Hoque}, \citenamefont {Karpiak},\ and\
  \citenamefont {Dash}}]{Khokhriakov_2020}%
  \BibitemOpen
  \bibfield  {author} {\bibinfo {author} {\bibfnamefont {D.}~\bibnamefont
  {Khokhriakov}}, \bibinfo {author} {\bibfnamefont {A.~M.}\ \bibnamefont
  {Hoque}}, \bibinfo {author} {\bibfnamefont {B.}~\bibnamefont {Karpiak}},\
  and\ \bibinfo {author} {\bibfnamefont {S.~P.}\ \bibnamefont {Dash}},\
  }\bibfield  {title} {\bibinfo {title} {Gate-tunable spin-galvanic effect in
  graphene-topological insulator van der waals heterostructures at room
  temperature},\ }\bibfield  {journal} {\bibinfo  {journal} {Nature
  Communications}\ }\textbf {\bibinfo {volume} {11}},\ \href
  {https://doi.org/10.1038/s41467-020-17481-1} {10.1038/s41467-020-17481-1}
  (\bibinfo {year} {2020})\BibitemShut {NoStop}%
\bibitem [{\citenamefont {Fonseca}\ \emph
  {et~al.}(2023{\natexlab{a}})\citenamefont {Fonseca}, \citenamefont
  {Pereira},\ and\ \citenamefont {Barbosa}}]{PhysRevB.107.155432}%
  \BibitemOpen
  \bibfield  {author} {\bibinfo {author} {\bibfnamefont {D.~B.}\ \bibnamefont
  {Fonseca}}, \bibinfo {author} {\bibfnamefont {L.~F.~C.}\ \bibnamefont
  {Pereira}},\ and\ \bibinfo {author} {\bibfnamefont {A.~L.}\ \bibnamefont
  {Barbosa}},\ }\bibfield  {title} {\bibinfo {title} {L{\'e}vy flight for
  electrons in graphene: Superdiffusive-to-diffusive transport transition},\
  }\href {https://doi.org/10.1103/PhysRevB.107.155432} {\bibfield  {journal}
  {\bibinfo  {journal} {Physical Review B}\ }\textbf {\bibinfo {volume}
  {107}},\ \bibinfo {pages} {155432} (\bibinfo {year}
  {2023}{\natexlab{a}})}\BibitemShut {NoStop}%
\bibitem [{\citenamefont {Fonseca}\ \emph {et~al.}(2024)\citenamefont
  {Fonseca}, \citenamefont {Barbosa},\ and\ \citenamefont
  {Pereira}}]{PhysRevB.110.075421}%
  \BibitemOpen
  \bibfield  {author} {\bibinfo {author} {\bibfnamefont {D.~B.}\ \bibnamefont
  {Fonseca}}, \bibinfo {author} {\bibfnamefont {A.~L.~R.}\ \bibnamefont
  {Barbosa}},\ and\ \bibinfo {author} {\bibfnamefont {L.~F.~C.}\ \bibnamefont
  {Pereira}},\ }\bibfield  {title} {\bibinfo {title} {L\'evy flight for
  electrons in graphene in the presence of regions with enhanced spin-orbit
  coupling},\ }\href {https://doi.org/10.1103/PhysRevB.110.075421} {\bibfield
  {journal} {\bibinfo  {journal} {Phys. Rev. B}\ }\textbf {\bibinfo {volume}
  {110}},\ \bibinfo {pages} {075421} (\bibinfo {year} {2024})}\BibitemShut
  {NoStop}%
\bibitem [{\citenamefont {Caridad}\ \emph {et~al.}(2016)\citenamefont
  {Caridad}, \citenamefont {Connaughton}, \citenamefont {Ott}, \citenamefont
  {Weber},\ and\ \citenamefont {Krsti\'c}}]{Caridad2016}%
  \BibitemOpen
  \bibfield  {author} {\bibinfo {author} {\bibfnamefont {J.~M.}\ \bibnamefont
  {Caridad}}, \bibinfo {author} {\bibfnamefont {S.}~\bibnamefont
  {Connaughton}}, \bibinfo {author} {\bibfnamefont {C.}~\bibnamefont {Ott}},
  \bibinfo {author} {\bibfnamefont {H.~B.}\ \bibnamefont {Weber}},\ and\
  \bibinfo {author} {\bibfnamefont {V.}~\bibnamefont {Krsti\'c}},\ }\bibfield
  {title} {\bibinfo {title} {An electrical analogy to mie scattering},\ }\href
  {https://doi.org/doi.org/10.1038/ncomms12894} {\bibfield  {journal} {\bibinfo
   {journal} {Nature Communications}\ }\textbf {\bibinfo {volume} {7}},\
  \bibinfo {pages} {12894} (\bibinfo {year} {2016})}\BibitemShut {NoStop}%
\bibitem [{\citenamefont {Gutiérrez}\ \emph {et~al.}(2016)\citenamefont
  {Gutiérrez}, \citenamefont {Brown}, \citenamefont {Kim}, \citenamefont
  {Park},\ and\ \citenamefont {Pasupathy}}]{Christopher2016}%
  \BibitemOpen
  \bibfield  {author} {\bibinfo {author} {\bibfnamefont {C.}~\bibnamefont
  {Gutiérrez}}, \bibinfo {author} {\bibfnamefont {L.}~\bibnamefont {Brown}},
  \bibinfo {author} {\bibfnamefont {C.-J.}\ \bibnamefont {Kim}}, \bibinfo
  {author} {\bibfnamefont {J.}~\bibnamefont {Park}},\ and\ \bibinfo {author}
  {\bibfnamefont {A.}~\bibnamefont {Pasupathy}},\ }\bibfield  {title} {\bibinfo
  {title} {Klein tunnelling and electron trapping in nanometre-scale graphene
  quantum dots},\ }\href {https://doi.org/10.1038/nphys3806} {\bibfield
  {journal} {\bibinfo  {journal} {Nature Physics}\ }\textbf {\bibinfo {volume}
  {12}} (\bibinfo {year} {2016})}\BibitemShut {NoStop}%
\bibitem [{\citenamefont {Fehske}\ \emph {et~al.}(2015)\citenamefont {Fehske},
  \citenamefont {Hager},\ and\ \citenamefont
  {Pieper}}]{https://doi.org/10.1002/pssb.201552119}%
  \BibitemOpen
  \bibfield  {author} {\bibinfo {author} {\bibfnamefont {H.}~\bibnamefont
  {Fehske}}, \bibinfo {author} {\bibfnamefont {G.}~\bibnamefont {Hager}},\ and\
  \bibinfo {author} {\bibfnamefont {A.}~\bibnamefont {Pieper}},\ }\bibfield
  {title} {\bibinfo {title} {Electron confinement in graphene with gate-defined
  quantum dots},\ }\href
  {https://doi.org/https://doi.org/10.1002/pssb.201552119} {\bibfield
  {journal} {\bibinfo  {journal} {physica status solidi (b)}\ }\textbf
  {\bibinfo {volume} {252}},\ \bibinfo {pages} {1868} (\bibinfo {year}
  {2015})},\ \Eprint
  {https://arxiv.org/abs/https://onlinelibrary.wiley.com/doi/pdf/10.1002/pssb.201552119}
  {https://onlinelibrary.wiley.com/doi/pdf/10.1002/pssb.201552119} \BibitemShut
  {NoStop}%
\bibitem [{\citenamefont {Sadrara}\ and\ \citenamefont
  {Miri}(2019)}]{PhysRevB.99.155432}%
  \BibitemOpen
  \bibfield  {author} {\bibinfo {author} {\bibfnamefont {M.}~\bibnamefont
  {Sadrara}}\ and\ \bibinfo {author} {\bibfnamefont {M.}~\bibnamefont {Miri}},\
  }\bibfield  {title} {\bibinfo {title} {Dirac electron scattering from a
  cluster of electrostatically defined quantum dots in graphene},\ }\href
  {https://doi.org/10.1103/PhysRevB.99.155432} {\bibfield  {journal} {\bibinfo
  {journal} {Phys. Rev. B}\ }\textbf {\bibinfo {volume} {99}},\ \bibinfo
  {pages} {155432} (\bibinfo {year} {2019})}\BibitemShut {NoStop}%
\bibitem [{\citenamefont {Safeer}\ \emph {et~al.}(2020)\citenamefont {Safeer},
  \citenamefont {Ingla-Aynés}, \citenamefont {Ontoso}, \citenamefont
  {Herling}, \citenamefont {Yan}, \citenamefont {Hueso},\ and\ \citenamefont
  {Casanova}}]{Safeer_2020}%
  \BibitemOpen
  \bibfield  {author} {\bibinfo {author} {\bibfnamefont {C.~K.}\ \bibnamefont
  {Safeer}}, \bibinfo {author} {\bibfnamefont {J.}~\bibnamefont
  {Ingla-Aynés}}, \bibinfo {author} {\bibfnamefont {N.}~\bibnamefont
  {Ontoso}}, \bibinfo {author} {\bibfnamefont {F.}~\bibnamefont {Herling}},
  \bibinfo {author} {\bibfnamefont {W.}~\bibnamefont {Yan}}, \bibinfo {author}
  {\bibfnamefont {L.~E.}\ \bibnamefont {Hueso}},\ and\ \bibinfo {author}
  {\bibfnamefont {F.}~\bibnamefont {Casanova}},\ }\bibfield  {title} {\bibinfo
  {title} {Spin hall effect in bilayer graphene combined with an insulator up
  to room temperature},\ }\href {https://doi.org/10.1021/acs.nanolett.0c01428}
  {\bibfield  {journal} {\bibinfo  {journal} {Nano Letters}\ }\textbf {\bibinfo
  {volume} {20}},\ \bibinfo {pages} {4573–4579} (\bibinfo {year}
  {2020})}\BibitemShut {NoStop}%
\bibitem [{\citenamefont {Yang}\ \emph
  {et~al.}(2023{\natexlab{b}})\citenamefont {Yang}, \citenamefont {Ormaza},
  \citenamefont {Chi}, \citenamefont {Dolan}, \citenamefont {Ingla-Aynés},
  \citenamefont {Safeer}, \citenamefont {Herling}, \citenamefont {Ontoso},
  \citenamefont {Gobbi}, \citenamefont {Martín-García}, \citenamefont
  {Schiller}, \citenamefont {Hueso},\ and\ \citenamefont
  {Casanova}}]{Yang_2023}%
  \BibitemOpen
  \bibfield  {author} {\bibinfo {author} {\bibfnamefont {H.}~\bibnamefont
  {Yang}}, \bibinfo {author} {\bibfnamefont {M.}~\bibnamefont {Ormaza}},
  \bibinfo {author} {\bibfnamefont {Z.}~\bibnamefont {Chi}}, \bibinfo {author}
  {\bibfnamefont {E.}~\bibnamefont {Dolan}}, \bibinfo {author} {\bibfnamefont
  {J.}~\bibnamefont {Ingla-Aynés}}, \bibinfo {author} {\bibfnamefont
  {C.}~\bibnamefont {Safeer}}, \bibinfo {author} {\bibfnamefont
  {F.}~\bibnamefont {Herling}}, \bibinfo {author} {\bibfnamefont
  {N.}~\bibnamefont {Ontoso}}, \bibinfo {author} {\bibfnamefont
  {M.}~\bibnamefont {Gobbi}}, \bibinfo {author} {\bibfnamefont
  {B.}~\bibnamefont {Martín-García}}, \bibinfo {author} {\bibfnamefont
  {F.}~\bibnamefont {Schiller}}, \bibinfo {author} {\bibfnamefont {L.~E.}\
  \bibnamefont {Hueso}},\ and\ \bibinfo {author} {\bibfnamefont
  {F.}~\bibnamefont {Casanova}},\ }\bibfield  {title} {\bibinfo {title}
  {Gate-tunable spin hall effect in an all-light-element heterostructure:
  Graphene with copper oxide},\ }\href
  {https://doi.org/10.1021/acs.nanolett.3c00687} {\bibfield  {journal}
  {\bibinfo  {journal} {Nano Letters}\ }\textbf {\bibinfo {volume} {23}},\
  \bibinfo {pages} {4406–4414} (\bibinfo {year}
  {2023}{\natexlab{b}})}\BibitemShut {NoStop}%
\bibitem [{\citenamefont {Nikoli\ifmmode~\acute{c}\else \'{c}\fi{}}\ \emph
  {et~al.}(2005)\citenamefont {Nikoli\ifmmode~\acute{c}\else \'{c}\fi{}},
  \citenamefont {Z\^arbo},\ and\ \citenamefont {Souma}}]{PhysRevB.72.075361}%
  \BibitemOpen
  \bibfield  {author} {\bibinfo {author} {\bibfnamefont {B.~K.}\ \bibnamefont
  {Nikoli\ifmmode~\acute{c}\else \'{c}\fi{}}}, \bibinfo {author} {\bibfnamefont
  {L.~P.}\ \bibnamefont {Z\^arbo}},\ and\ \bibinfo {author} {\bibfnamefont
  {S.}~\bibnamefont {Souma}},\ }\bibfield  {title} {\bibinfo {title}
  {Mesoscopic spin hall effect in multiprobe ballistic spin-orbit-coupled
  semiconductor bridges},\ }\href {https://doi.org/10.1103/PhysRevB.72.075361}
  {\bibfield  {journal} {\bibinfo  {journal} {Phys. Rev. B}\ }\textbf {\bibinfo
  {volume} {72}},\ \bibinfo {pages} {075361} (\bibinfo {year}
  {2005})}\BibitemShut {NoStop}%
\bibitem [{\citenamefont {Bardarson}\ \emph {et~al.}(2007)\citenamefont
  {Bardarson}, \citenamefont {Adagideli},\ and\ \citenamefont
  {Jacquod}}]{PhysRevLett.98.196601}%
  \BibitemOpen
  \bibfield  {author} {\bibinfo {author} {\bibfnamefont {J.~H.}\ \bibnamefont
  {Bardarson}}, \bibinfo {author} {\bibfnamefont {i.~d.~I.}\ \bibnamefont
  {Adagideli}},\ and\ \bibinfo {author} {\bibfnamefont {P.}~\bibnamefont
  {Jacquod}},\ }\bibfield  {title} {\bibinfo {title} {Mesoscopic spin hall
  effect},\ }\href {https://doi.org/10.1103/PhysRevLett.98.196601} {\bibfield
  {journal} {\bibinfo  {journal} {Phys. Rev. Lett.}\ }\textbf {\bibinfo
  {volume} {98}},\ \bibinfo {pages} {196601} (\bibinfo {year}
  {2007})}\BibitemShut {NoStop}%
\bibitem [{\citenamefont {Santana}\ \emph {et~al.}(2020)\citenamefont
  {Santana}, \citenamefont {da~Silva}, \citenamefont {Vasconcelos},
  \citenamefont {Ramos},\ and\ \citenamefont {Barbosa}}]{PhysRevB.102.041107}%
  \BibitemOpen
  \bibfield  {author} {\bibinfo {author} {\bibfnamefont {F.~A.~F.}\
  \bibnamefont {Santana}}, \bibinfo {author} {\bibfnamefont {J.~M.}\
  \bibnamefont {da~Silva}}, \bibinfo {author} {\bibfnamefont {T.~C.}\
  \bibnamefont {Vasconcelos}}, \bibinfo {author} {\bibfnamefont {J.~G. G.~S.}\
  \bibnamefont {Ramos}},\ and\ \bibinfo {author} {\bibfnamefont {A.~L.~R.}\
  \bibnamefont {Barbosa}},\ }\bibfield  {title} {\bibinfo {title} {Spin hall
  angle fluctuations in a device with disorder},\ }\href
  {https://doi.org/10.1103/PhysRevB.102.041107} {\bibfield  {journal} {\bibinfo
   {journal} {Phys. Rev. B}\ }\textbf {\bibinfo {volume} {102}},\ \bibinfo
  {pages} {041107} (\bibinfo {year} {2020})}\BibitemShut {NoStop}%
\bibitem [{\citenamefont {da~Silva}\ \emph {et~al.}(2022)\citenamefont
  {da~Silva}, \citenamefont {Santana}, \citenamefont {Ramos},\ and\
  \citenamefont {Barbosa}}]{10.1063/5.0107212}%
  \BibitemOpen
  \bibfield  {author} {\bibinfo {author} {\bibfnamefont {J.~M.}\ \bibnamefont
  {da~Silva}}, \bibinfo {author} {\bibfnamefont {F.~A.~F.}\ \bibnamefont
  {Santana}}, \bibinfo {author} {\bibfnamefont {J.~G. G.~S.}\ \bibnamefont
  {Ramos}},\ and\ \bibinfo {author} {\bibfnamefont {A.~L.~R.}\ \bibnamefont
  {Barbosa}},\ }\bibfield  {title} {\bibinfo {title} {{Spin Hall angle in
  single-layer graphene}},\ }\href {https://doi.org/10.1063/5.0107212}
  {\bibfield  {journal} {\bibinfo  {journal} {Journal of Applied Physics}\
  }\textbf {\bibinfo {volume} {132}},\ \bibinfo {pages} {183901} (\bibinfo
  {year} {2022})},\ \Eprint
  {https://arxiv.org/abs/https://pubs.aip.org/aip/jap/article-pdf/doi/10.1063/5.0107212/16516946/183901\_1\_online.pdf}
  {https://pubs.aip.org/aip/jap/article-pdf/doi/10.1063/5.0107212/16516946/183901\_1\_online.pdf}
  \BibitemShut {NoStop}%
\bibitem [{\citenamefont {Fonseca}\ \emph
  {et~al.}(2023{\natexlab{b}})\citenamefont {Fonseca}, \citenamefont
  {Pereira},\ and\ \citenamefont {Barbosa}}]{PhysRevB.108.245105}%
  \BibitemOpen
  \bibfield  {author} {\bibinfo {author} {\bibfnamefont {D.~B.}\ \bibnamefont
  {Fonseca}}, \bibinfo {author} {\bibfnamefont {L.~L.~A.}\ \bibnamefont
  {Pereira}},\ and\ \bibinfo {author} {\bibfnamefont {A.~L.~R.}\ \bibnamefont
  {Barbosa}},\ }\bibfield  {title} {\bibinfo {title} {Orbital hall effect in
  mesoscopic devices},\ }\href {https://doi.org/10.1103/PhysRevB.108.245105}
  {\bibfield  {journal} {\bibinfo  {journal} {Phys. Rev. B}\ }\textbf {\bibinfo
  {volume} {108}},\ \bibinfo {pages} {245105} (\bibinfo {year}
  {2023}{\natexlab{b}})}\BibitemShut {NoStop}%
\bibitem [{\citenamefont {Barbosa}\ \emph {et~al.}(2024)\citenamefont
  {Barbosa}, \citenamefont {Canonico}, \citenamefont {Garc\'{\i}a},\ and\
  \citenamefont {Rappoport}}]{PhysRevB.110.085412}%
  \BibitemOpen
  \bibfield  {author} {\bibinfo {author} {\bibfnamefont {A.~L.~R.}\
  \bibnamefont {Barbosa}}, \bibinfo {author} {\bibfnamefont {L.~M.}\
  \bibnamefont {Canonico}}, \bibinfo {author} {\bibfnamefont {J.~H.}\
  \bibnamefont {Garc\'{\i}a}},\ and\ \bibinfo {author} {\bibfnamefont {T.~G.}\
  \bibnamefont {Rappoport}},\ }\bibfield  {title} {\bibinfo {title} {Orbital
  hall effect and topology on a two-dimensional triangular lattice: From bulk
  to edge},\ }\href {https://doi.org/10.1103/PhysRevB.110.085412} {\bibfield
  {journal} {\bibinfo  {journal} {Phys. Rev. B}\ }\textbf {\bibinfo {volume}
  {110}},\ \bibinfo {pages} {085412} (\bibinfo {year} {2024})}\BibitemShut
  {NoStop}%
\bibitem [{\citenamefont {Barbosa}\ \emph {et~al.}(2025)\citenamefont
  {Barbosa}, \citenamefont {Lee},\ and\ \citenamefont
  {Rappoport}}]{barbosa2025extrinsicorbitalhalleffect}%
  \BibitemOpen
  \bibfield  {author} {\bibinfo {author} {\bibfnamefont {A.~L.~R.}\
  \bibnamefont {Barbosa}}, \bibinfo {author} {\bibfnamefont {H.-W.}\
  \bibnamefont {Lee}},\ and\ \bibinfo {author} {\bibfnamefont {T.~G.}\
  \bibnamefont {Rappoport}},\ }\href {https://arxiv.org/abs/2507.01941}
  {\bibinfo {title} {Extrinsic orbital hall effect and orbital relaxation in
  mesoscopic devices}} (\bibinfo {year} {2025}),\ \Eprint
  {https://arxiv.org/abs/2507.01941} {arXiv:2507.01941 [cond-mat.mes-hall]}
  \BibitemShut {NoStop}%
\bibitem [{\citenamefont {Groth}\ \emph {et~al.}(2014)\citenamefont {Groth},
  \citenamefont {Wimmer}, \citenamefont {Akhmerov},\ and\ \citenamefont
  {Waintal}}]{kwant}%
  \BibitemOpen
  \bibfield  {author} {\bibinfo {author} {\bibfnamefont {C.~W.}\ \bibnamefont
  {Groth}}, \bibinfo {author} {\bibfnamefont {M.}~\bibnamefont {Wimmer}},
  \bibinfo {author} {\bibfnamefont {A.~R.}\ \bibnamefont {Akhmerov}},\ and\
  \bibinfo {author} {\bibfnamefont {X.}~\bibnamefont {Waintal}},\ }\bibfield
  {title} {\bibinfo {title} {Kwant: a software package for quantum transport},\
  }\href {https://doi.org/10.1088/1367-2630/16/6/063065} {\bibfield  {journal}
  {\bibinfo  {journal} {New Journal of Physics}\ }\textbf {\bibinfo {volume}
  {16}},\ \bibinfo {pages} {063065} (\bibinfo {year} {2014})}\BibitemShut
  {NoStop}%
\bibitem [{\citenamefont {Cysne}\ \emph
  {et~al.}(2018{\natexlab{a}})\citenamefont {Cysne}, \citenamefont {Ferreira},\
  and\ \citenamefont {Rappoport}}]{PhysRevB.98.045407}%
  \BibitemOpen
  \bibfield  {author} {\bibinfo {author} {\bibfnamefont {T.~P.}\ \bibnamefont
  {Cysne}}, \bibinfo {author} {\bibfnamefont {A.}~\bibnamefont {Ferreira}},\
  and\ \bibinfo {author} {\bibfnamefont {T.~G.}\ \bibnamefont {Rappoport}},\
  }\bibfield  {title} {\bibinfo {title} {Crystal-field effects in graphene with
  interface-induced spin-orbit coupling},\ }\href
  {https://doi.org/10.1103/PhysRevB.98.045407} {\bibfield  {journal} {\bibinfo
  {journal} {Phys. Rev. B}\ }\textbf {\bibinfo {volume} {98}},\ \bibinfo
  {pages} {045407} (\bibinfo {year} {2018}{\natexlab{a}})}\BibitemShut
  {NoStop}%
\bibitem [{\citenamefont {Cysne}\ \emph
  {et~al.}(2018{\natexlab{b}})\citenamefont {Cysne}, \citenamefont {Garcia},
  \citenamefont {Rocha},\ and\ \citenamefont {Rappoport}}]{PhysRevB.97.085413}%
  \BibitemOpen
  \bibfield  {author} {\bibinfo {author} {\bibfnamefont {T.~P.}\ \bibnamefont
  {Cysne}}, \bibinfo {author} {\bibfnamefont {J.~H.}\ \bibnamefont {Garcia}},
  \bibinfo {author} {\bibfnamefont {A.~R.}\ \bibnamefont {Rocha}},\ and\
  \bibinfo {author} {\bibfnamefont {T.~G.}\ \bibnamefont {Rappoport}},\
  }\bibfield  {title} {\bibinfo {title} {Quantum hall effect in graphene with
  interface-induced spin-orbit coupling},\ }\href
  {https://doi.org/10.1103/PhysRevB.97.085413} {\bibfield  {journal} {\bibinfo
  {journal} {Phys. Rev. B}\ }\textbf {\bibinfo {volume} {97}},\ \bibinfo
  {pages} {085413} (\bibinfo {year} {2018}{\natexlab{b}})}\BibitemShut
  {NoStop}%
\bibitem [{\citenamefont {Sousa-J\'unior}\ \emph {et~al.}(2025)\citenamefont
  {Sousa-J\'unior}, \citenamefont {Ferraz}, \citenamefont {de~Lima},\ and\
  \citenamefont {Cysne}}]{PhysRevB.111.035411}%
  \BibitemOpen
  \bibfield  {author} {\bibinfo {author} {\bibfnamefont {S.~a. d.~A.}\
  \bibnamefont {Sousa-J\'unior}}, \bibinfo {author} {\bibfnamefont {M.~V.
  d.~S.}\ \bibnamefont {Ferraz}}, \bibinfo {author} {\bibfnamefont {J.~P.}\
  \bibnamefont {de~Lima}},\ and\ \bibinfo {author} {\bibfnamefont {T.~P.}\
  \bibnamefont {Cysne}},\ }\bibfield  {title} {\bibinfo {title} {Topological
  characterization of modified kane-mele-rashba models via local spin chern
  marker},\ }\href {https://doi.org/10.1103/PhysRevB.111.035411} {\bibfield
  {journal} {\bibinfo  {journal} {Phys. Rev. B}\ }\textbf {\bibinfo {volume}
  {111}},\ \bibinfo {pages} {035411} (\bibinfo {year} {2025})}\BibitemShut
  {NoStop}%
\bibitem [{\citenamefont {Castro~Neto}\ \emph {et~al.}(2009)\citenamefont
  {Castro~Neto}, \citenamefont {Guinea}, \citenamefont {Peres}, \citenamefont
  {Novoselov},\ and\ \citenamefont {Geim}}]{RevModPhys.81.109}%
  \BibitemOpen
  \bibfield  {author} {\bibinfo {author} {\bibfnamefont {A.~H.}\ \bibnamefont
  {Castro~Neto}}, \bibinfo {author} {\bibfnamefont {F.}~\bibnamefont {Guinea}},
  \bibinfo {author} {\bibfnamefont {N.~M.~R.}\ \bibnamefont {Peres}}, \bibinfo
  {author} {\bibfnamefont {K.~S.}\ \bibnamefont {Novoselov}},\ and\ \bibinfo
  {author} {\bibfnamefont {A.~K.}\ \bibnamefont {Geim}},\ }\bibfield  {title}
  {\bibinfo {title} {The electronic properties of graphene},\ }\href
  {https://doi.org/10.1103/RevModPhys.81.109} {\bibfield  {journal} {\bibinfo
  {journal} {Rev. Mod. Phys.}\ }\textbf {\bibinfo {volume} {81}},\ \bibinfo
  {pages} {109} (\bibinfo {year} {2009})}\BibitemShut {NoStop}%
\end{thebibliography}%


\end{document}